\documentclass{ws-ijmpd}
\usepackage[utf8]{inputenc}
\usepackage[super,compress]{cite}
\usepackage{graphicx}
\usepackage{latexsym}
\usepackage{amssymb, amsmath}
\usepackage{fourier, bm, physics, color, xcolor}
\usepackage[colorlinks,citecolor=blue,urlcolor=blue,linkcolor=blue]{hyperref}
\usepackage{orcidlink}
\usepackage{eucal}

\begin{document}

\markboth{Luca Visinelli}
{Boson stars: a review}
\catchline{}{}{}{}{}

\title{\Large BOSON STARS AND OSCILLATONS: A REVIEW}

\author{\Large LUCA VISINELLI}
\vspace{0.1cm}
\address{\normalsize	 Tsung-Dao Lee Institute (TDLI), Shanghai Jiao Tong University, 200240 Shanghai, China\\ 
\vspace{0.1cm}
and \\
\vspace{0.1cm}
INFN, Laboratori Nazionali di Frascati, C.P. 13, 100044 Frascati, Italy\\
\vspace{0.3cm}
\href{mailto:luca.visinelli@sjtu.edu.cn}{luca.visinelli@sjtu.edu.cn}}

\maketitle

\begin{abstract}
Compact objects occupy a pivotal role in the exploration of Nature. The interest spans from the role of compact objects in astrophysics to their detection through various methods (gravitational waves interferometry, microlensing, imaging). While the existence of compact objects made of fermions (neutron stars and white dwarfs) has been assessed, a parallel search for localized solitons made of bosons is ongoing, stemming from Wheeler's original proposal of electromagnetic ``geons''. Boson fields can clump up and form compact objects such as boson stars (for complex scalar fields), oscillons and oscillatons (for real scalar fields), or Proca stars (for massive vector fields), which can show up in searches as black hole mimickers, dark matter sources, and a variety of other phenomena. I review some crucial properties of these bosonic systems, including recent progress in the field.
\end{abstract}


\tableofcontents

\section{Introduction}

The idea that particles might collectively arrange to form stable macroscopic objects has long been explored. In 1955, John Wheeler proposed a solution of the classical electromagnetic field in the context of General Relativity (GR)~\cite{Wheeler:1955zz}. These configurations, called ``geons'', turned out to be unstable against perturbations~\cite{Regge:1957td, Power:1957zz}. However, when a complex scalar field is considered in place of the electromagnetic field, the equation of motion for the complex scalar field (the Klein-Gordon equation) admits localized and stable geon-like solutions known as boson stars (BSs)~\cite{Kaup:1968zz}.

The interest in BSs has experienced a resurgence following the confirmation of the Higgs boson's existence~\cite{CMS:2012qbp, ATLAS:2012yve}, which shows that fundamental bosons are indeed part of Nature. Although the Higgs boson is unstable and rapidly decays to W and Z bosons, other fundamental complex scalar fields could generally arise in theories that extend the Standard Model (SM) of particle physics. For example, the Peccei-Quinn mechanism~\cite{Peccei:1977ur, Peccei:1977hh} is one of the best motivated solutions to the strong-CP problem of QCD and predicts the QCD axion~\cite{Weinberg:1977ma, Wilczek:1977pj}; the simplest models of inflation that extends the standard cosmological model requires a scalar inflaton field; supersymmetric extensions of the SM predict scalar fields of masses above the electroweak scale~\cite{Roszkowski:1993by, Martin:1997ns}; string theory predicts the existence of several moduli fields that encode the properties of extra dimensions and axion-like particles~\cite{Arvanitaki:2009fg}. There are at least two more core reasons to study BSs. First of all, a BS serves as a laboratory to assess the properties of compact self-gravitating objects. In addition, due to their mass range and the stability, it is conceivable to postulate that BSs could be detected through some astrophysical method, and could even serve as the dark matter in the Universe.

The list of previous reviews on the subject is quite long and indicates the rapid progresses in the field. Jetzer~\cite{Jetzer:1991jr} and Liddle \& Madsen~\cite{Liddle:1992fmk} focus on the importance of BS in astrophysics and the mechanisms of formation, respectively. Lee \& Pang~\cite{Lee:1991ax} generally review the crucial aspects of topological and non-topological solitons. Straumann discusses the similarities and differences of fermion and boson stars~\cite{Straumann:1991pt}. Mielke \& Schunck~\cite{Schunck:1999pm, Schunck:2003kk} discuss in depth on the possibility to detect BSs in astrophysics. Liebling \& Palenzuela~\cite{Liebling:2012fv} focus on the various solutions and on the astrophysical signatures of BSs. The review of Krippendorf, Muia \& Quevedo~\cite{Krippendorf:2018tei} focuses on all possible compact objects that can form within the theory of string compactification. Excellent PhD theses on the numerical treatment of BS solutions are given by Helfer~\cite{Helfer:2020gui} and by Fodor~\cite{Fodor:2019ftc}.

This review has come to existence in light of the recent and prolific advance in the theory, phenomenology, and numerical simulations of the formation and evolution of compact objects made of bosons. Since the most recent reviews on the subject appeared, many concepts have been further scrutinized such as the study of Proca stars and solitons in modified gravity theory, a list of numerical studies on the gravitational cooling and on the merging of compact objects, the emergence and maturity of astrophysical phenomena such as gravitational wave searches and lensing, the possibility that compact objects explain the observed dark matter abundance. Here, we generically refer to a {\it non-topological soliton} to describe a long-lived and spatially localized compact object whose stability is granted by a conserved charge, a (mini- or massive) {\it boson star} for a non-topological soliton which is supported by gravity, a {\it Q-ball} for a non-topological soliton supported by self-interactions, a {\it pseudo-soliton} for a metastable configuration made out of a real scalar field for which a conserved charge does not exist, an {\it oscillaton} if the pseudo-soliton is made out of a real scalar field and it is supported by gravity, and an {\it oscillon} if the pseudo-soliton is supported by self-interactions. In this view, a pseudo-soliton is not necessarily an oscillon.

This review is organized as follows. In Sec.~\ref{sec:BS} we describe some general considerations about the stability of BSs. In Sec.~\ref{sec:complex}, we derive the equations describing a stable BS from the relativistic action for a complex scalar field, focusing on specific solutions for the case of mini-BSs~(\ref{sec:miniBS}), massive BSs~(\ref{sec:selfinteracting}), and non-gravitational solitonic solutions~(\ref{sec:nongravitationalBS}). In Sec.~\ref{sec:real} we introduce stable solutions for a real scalar field, including axion stars~(\ref{sec:axionstars}). We discuss rotating solutions in Sec.~\ref{sec:rotatingBS}. More speculative configurations are presented in Sec.~\ref{sec:speculative}. Macroscopic Bose-Einstein condensation is discussed in Sec.~\ref{sec:BEC}. The formation of BSs is reviewed in Sec.~\ref{sec:formation}. Stability of BSs against perturbations is discussed in Sec.~\ref{sec:stability}. Sec.~\ref{sec:GW} explores the detection of gravitational waves from BS compact binaries. The role of BSs as the dark matter is presented in Sec.~\ref{sec:DM}, and other signatures are discussed in Sec.~\ref{sec:othersearches}.

We work in natural units $c = \hbar = 1$ and we adopt the metric signature $(-, +, +, +)$. In these units, the Planck mass is $m_{\rm Pl} = 1/\sqrt{G}$, and we also introduce the reduced Planck mass $M_{\rm Pl} = (8\pi G)^{-1/2} = m_{\rm Pl}/\sqrt{8\pi}$. A brief list of acronyms and notation used is given in Table~\ref{tabnotation}.

\begin{table}[ht]
\centering
\tbl{List of conventions and acronyms used in this review.}{
\begin{tabular}{|l|l|}
\hline
\hspace{0.5cm} Greek small letters $\alpha, \beta,$... & \hspace{0.5cm} Spacetime coordinates indices\\
\hspace{0.5cm} Latin small letters $i,j,k$... & \hspace{0.5cm} Space coordinates indices\\
\hspace{0.5cm} $g_{\alpha\beta}$ & \hspace{0.5cm} Metric tensor \\
\hspace{0.5cm} $\left(-, +, +, +\right)$ & \hspace{0.5cm} Metric signature \\
\hspace{0.5cm} $m_{\rm Pl} \equiv 1/\sqrt{G}$; $M_{\rm Pl} \equiv 1/\sqrt{8\pi G}$ & \hspace{0.5cm} Planck mass; Reduced Planck mass\\
\hspace{0.5cm} $\mu$ & \hspace{0.5cm} Bare boson mass\\
\hspace{0.5cm} $\Phi$; $\phi$ & \hspace{0.5cm} Complex scalar field; Real scalar field\\
\hspace{0.5cm} $\psi$ & \hspace{0.5cm} Non-relativistic wave function\\
\hspace{0.5cm} BS & \hspace{0.5cm} Boson star\\
\hspace{0.5cm} BH & \hspace{0.5cm} Black hole\\
\hspace{0.5cm} NS & \hspace{0.5cm} Neutron star\\
\hspace{0.5cm} EKG & \hspace{0.5cm} Einstein-Klein-Gordon Eqs.~\eqref{eq:Einstein} and~\eqref{eq:KGE}\\
\hline
\end{tabular}}
\label{tabnotation}
\end{table}

\section{Boson Stars}
\label{sec:BS}

Boson stars (BS) are soliton solutions of the Klein-Gordon equation describing a classical complex scalar field $\Phi$ coupled to gravity. The self-gravitational energy of the boson field is sourced by the spatial gradients and time derivative of the field itself. It is the dispersive nature of the Klein-Gordon equation that provides the sufficient pressure to balance the gravitational field sourced by the self-gravity of the bosons. Such a wave-like behavior is ultimately related to Heisenberg uncertainty principle, which states that a particle of mass $\mu$ confined within an object of size $2R$ (where $R$ is the radius of the star) possesses a velocity $v \sim (2R\mu)^{-1}$. The total kinetic energy is then
\begin{equation}
	K = N\frac{\mu v^2}{2} \sim \frac{N}{8\mu R^2}\,,
\end{equation}
where $N$ is the total number of bosons in the BS. As we discuss in Sec.~\ref{sec:complex}, $N$ is a conserved quantity for a complex scalar field due to the U(1) symmetry that the field enjoys. The self-gravity potential energy of the configuration is $U \sim -(3/5)GM^2/R$, where $M \approx N\mu$ is the total mass of the star.\footnote{\,In this approximation, we are neglecting the binding energy of the star, which is crucial when performing the actual computation of the configuration.} The total energy is minimized at the configuration
\begin{equation}
	\label{eq:massradiusrelation}
	R_{\rm BS} \sim \frac{\alpha_k}{2G\mu^2M}\,,
\end{equation}
where $\alpha_k = 5/6$ from the estimate given. The formula in Eq.~\eqref{eq:massradiusrelation} is in general not satisfied by a BS, because the radial and transverse pressure terms are not equal. However, in the cases where an isotropic pressure can be assumed, the behaviour in Eq.~\eqref{eq:massradiusrelation} matches the numerical computation with a different values for $\alpha_k \approx 9.9$~\cite{Ruffini:1969qy}.

For the reason above, the BS is called a ``star'' not because it shines, but because it consists of an equilibrium configuration between the self-gravity and a pressure term. While for ordinary stars the internal pressure is provided by nuclear fusion reactions in the stellar core, for BSs it is Heisenberg uncertainty principle that provides the required ``quantum pressure'' term (see Sec.~\ref{sec:BEC} for a discussion). Both stars and BSs possess a negative heat capacity, in the sense that the average kinetic energy of each constituent increases when the system loses energy~\cite{Wallace:2009dc}.

The argument above also predicts a maximal value of the BS mass. In fact, given Eq.~\eqref{eq:massradiusrelation} a BS would shrink the more massive it becomes. However, there is an intrinsic limit of the radius, the Schwarzschild radius $R_S = 2GM$, below which the solution is no longer described by a BS and the system collapses into a black hole (BH). This corresponds to a mass
\begin{equation}
	\label{eq:maxmass}
	M_{\rm max} \propto m_{\rm Pl}^2/\mu\,,
\end{equation}
where $m_{\rm Pl}^2 = 1/G$. Indeed, in Sec.~\ref{sec:miniBS} we show with an explicit computation that a maximal mass exists for the simplest case of a free complex field, with the analogous of Eq.~\eqref{eq:maxmass} being known in the literature as the ``Kaup mass''~\cite{Kaup:1968zz, Breit:1983nr}. In this case, the occupation number $N_{\rm BS} \approx (m_{\rm Pl}/\mu)^2$ for a BS of nearly critical mass is much smaller than the corresponding value obtained for a fermion star, $N_{\rm FS} \sim (m_{\rm Pl}/\mu)^3$, hence these boson structures have been named ``mini-boson stars'' in the literature. The inverse proportionality between $M_{\rm max}$ and $\mu$ holds in the presence of a self-interacting potential as well, although the magnitude of the maximal mass can differ greatly from the free-field scenario and can even match the results obtained for the fermionic systems.

\section{Complex scalar field}
\label{sec:complex}

A complex scalar field $\Phi$ evolving in the GR framework is described by the Einstein-Klein-Gordon (EKG) action,
\begin{equation}
	\label{eq:action}
	\mathcal{S} = \int \mathrm{d}^4x\sqrt{-g} \left[\frac{\mathcal{R}}{16\pi G} - \nabla^\alpha\bar\Phi\nabla_\alpha\Phi - V(|\Phi|^2)\right]\,,
\end{equation}
where $\bar\Phi$ is the complex conjugate of $\Phi$, $V(|\Phi|^2)$ is the bosonic potential, $\mathcal{R}$ is the Ricci scalar, and $g$ is the determinant of the metric tensor $g^{\alpha\beta}$. The variation of the action with respect to the metric leads to the Einstein equations
\begin{equation}
	\label{eq:Einstein}
	\mathcal{R}_{\alpha\beta} - \frac{1}{2}g_{\alpha\beta}\mathcal{R} = 8\pi G \,T_{\alpha\beta}^\Phi\,
\end{equation}
where $\mathcal{R}_{\alpha\beta}$ is the Ricci tensor and the energy-momentum tensor of the boson field is
\begin{equation}
	\label{eq:energymomentum}
	T_{\alpha\beta}^\Phi = \nabla_\alpha\bar\Phi \nabla_\beta\Phi + \nabla_\beta\bar\Phi \nabla_\alpha\Phi - g_{\alpha\beta}\left(\nabla^\gamma\bar\Phi\nabla_\gamma\Phi + V(|\Phi|^2)\right)\,.
\end{equation}
In the following, we also use the equivalent formulation of Einstein's equations,
\begin{equation}
	\label{eq:Einstein1}
	\mathcal{R}_{\alpha\beta} = 8\pi G \,\left(T_{\alpha\beta}^\Phi - \frac{1}{2}g_{\alpha\beta}T^\Phi\right)\,,
\end{equation}
where $T^\Phi$ is the trace of $T^\Phi_{\alpha\beta}$. The variation of the action with respect to $\bar\Phi$ leads to the Klein-Gordon equation,
\begin{equation}
	\label{eq:KGE}
	\nabla^\alpha\nabla_\alpha\Phi = \frac{\mathrm{d}V(|\Phi|^2)}{\mathrm{d}|\Phi|^2}\Phi\,.
\end{equation}
In the following, we refer to the system of Eqs.~\eqref{eq:Einstein} and~\eqref{eq:KGE} as EKG.

The action Eq.~\eqref{eq:action} is invariant under the U(1) symmetry $\Phi \to \Phi\,e^{i\tau}$ for some angle $\tau$. According to Noether's theorem, this implies the existence of a conserved current
\begin{equation}
	\label{eq:noether}
	J_\mu = \frac{i}{2}\left(\bar\Phi \nabla_\mu \Phi - (\nabla_\mu \bar\Phi)\Phi \right)\,,
\end{equation}
so that the conservation law $\nabla_\mu J^\mu = 0$ is assured. The Noether charge, corresponding to the total number of bosons in the BS, is then
\begin{equation}
	\label{eq:noethercharge}
	N = \int \mathrm{d}^3x \sqrt{-g} g^{0\mu} J_\mu \,.
\end{equation}
A stable bound configuration such as a BS, in which the stability is granted by the existence of a conserved charge and whose boundary condition at infinity is the vacuum state, is referred to as a non-topological soliton~\cite{Lee:1991ax}. This solution differs from the {\it topological} soliton, for which an additional conservation law is not required and whose boundary condition at infinity is topologically different from the vacuum~\cite{tHooft:1974kcl, Polyakov:1974ek}. Promoting the U(1) symmetry to a local symmetry leads to charged BSs~\cite{Jetzer:1989av, Jetzer:1989us, Pugliese:2013gsa, Delgado:2016jxq, Kumar:2017zms, Kumar:2019dbi, Collodel:2019ohy}, whose collapse could lead to charged BHs~\cite{Jetzer:1992tog, Torres:2000dw, Kleihaus:2009kr}. The mass of the BS is provided by the Tolman mass formula~\cite{Tolman:1930zz, Papaetrou:1947ib, Landau:1975pou}
\begin{equation}
	\label{eq:tolmanmass}
	M = \int \mathrm{d}^3x \sqrt{-g} \left[2{T_0}^0-{T_\mu}^\mu\right] \approx \int \mathrm{d}^3x \sqrt{-g} \rho\,,
\end{equation}
where the last approximation is valid in the non-relativistic limit ${T_\mu}^\mu \approx {T_0}^0 \approx \rho$.

\subsection{Decomposing the solution}
\label{eq:decomposing}

A boson ``star'' is defined as a field configuration that is localized in space. A theorem due to Derrick~\cite{Derrick:1964} states that the Klein-Gordon equation does not admit a stable and localized time-independent solution. The physical argument on the impossibility to realize such a solution relies on the fact that shrinking a non-zero real scalar field configuration effectively reduces its total energy~\cite{Friedberg:1986tp}. Since the stress-energy tensor depends on the absolute value of the scalar field, Derrick's theorem can be circumvented by relaxing the assumption of time-independence by considering a time-periodic field, while retaining a time-independent gravitational field.\footnote{\,Another method to circumvent Derrick's theorem consists on invoking two or more interacting fields, such as the 't Hooft-Polyakov monopole or the Nielsen-Olesen vortices~\cite{Rajaraman:1982is}.}

We now present a solution to the set of equations in the case of spherical symmetry, adopting a harmonic ansatz for the complex scalar field which is decomposed in terms of a real radial function $\phi$ as
\begin{equation}
	\label{eq:ansatz}
	\Phi(r,t) = \phi(r)\,e^{-i\omega t}\,,
\end{equation}
where $\omega$ is a constant. In principle, the complex field can be written as a sum of real and imaginary parts as $\phi(r) = \phi_R(r) + i \phi_I(r)$. However, the two components follow the same equation so we treat $\phi(r)$ as a real function.

We restrict to the case of spherical symmetry to discuss the family of solutions for EKG. For a spherically symmetric BS, the metric in polar coordinates is of the form~\cite{Kaup:1968zz, Ruffini:1969qy}
\begin{equation}
	\label{eq:metric}
	\mathrm{d}s^2 = -e^{v}\, \mathrm{d}t^2 + e^{u}\, \mathrm{d}r^2 + r^2\left(\mathrm{d}\vartheta^2 + \sin^2\vartheta\mathrm{d}\varphi^2\right)\,,
\end{equation}
with the two functions $u = u(r)$ and $v=v(r)$. 

 With the ansatz in Eqs.~\eqref{eq:ansatz}-\eqref{eq:metric}, the EKG set of coupled Eqs.~\eqref{eq:Einstein}-\eqref{eq:KGE} reads
\begin{eqnarray}
	e^{-u}\left(\frac{u'}{r} - \frac{1}{r^2}\right) + \frac{1}{r^2} &=& 8\pi G\,\rho_\phi\,,\label{eq:u0}\\
	e^{-u}\left(\frac{v'}{r} + \frac{1}{r^2}\right) - \frac{1}{r^2} &=& 8\pi G \,p_{\rm rad}\,,\label{eq:v0}\\
	\phi'' + \left(\frac{2}{r} + \frac{v'-u'}{2}\right)\phi'(r) &=& e^{u}\left(\frac{\mathrm{d}V(\phi^2)}{\mathrm{d}\phi^2} -e^{-v}\omega^2\right)\phi\,,\label{eq:phi0}
\end{eqnarray}
where a prime denotes the differentiation with respect to the radial coordinate $r$. The effective density, radial pressure, and tangential pressure are given respectively by
\begin{eqnarray}
	\rho_\phi &=& e^{-v}\,\omega^2\,\phi^2 + e^{-u}(\phi')^2 + V(\phi^2) \,,\\
	p_{\rm rad} &=& e^{-v}\,\omega^2\,\phi^2 + e^{-u}(\phi')^2 - V(\phi^2) \,,\\
	p_{\rm tan} &=& e^{-v}\,\omega^2\,\phi^2 - e^{-u}(\phi')^2 - V(\phi^2) \,.
\end{eqnarray}
Since the radial and tangential pressure terms differ because of the different sign in $(\phi')^2$, the energy-momentum tensor is not generally isotropic for a BS and the fluid approximation is not valid. For this, the Tolman-Oppenheimer-Volkoff approach does not apply.

The coordinate $u(r)$ is related to mass conservation: in fact, once defined
\begin{equation}
	\label{eq:enclmass}
	e^{-u} = 1 - 2\mathcal{M}(r)/r\,,
\end{equation}
where $\mathcal{M}(r)$ is the mass enclosed within radius $r$, Eq.~\eqref{eq:u0} can be cast in the form
\begin{equation}
	\frac{\mathrm{d}\mathcal{M}(r)}{\mathrm{d}r} = 4\pi r^2\,\rho\,.
\end{equation}
The total mass of the BS is defined as $M = \lim\limits_{r\to+\infty}\mathcal{M}(r)$~\cite{Arnowitt:1959ah}.

\subsection{Mini-boson star}
\label{sec:miniBS}

The simplest potential to consider is of the quadratic form,\footnote{\,It is possible to obtain localized solutions for a set of massless fields minimally coupled to gravity~\cite{Hawley:2002zn}.}
\begin{equation}
	\label{eq:quadraticV}
	V(|\Phi|^2) = \mu^2|\Phi|^2\,,
\end{equation}
where $\mu$ is a constant quantity that represents the mass of the complex scalar field. We solve the system of Eqs.~\eqref{eq:u0}-\eqref{eq:phi0} for the case of this quadratic potential, rescaling the radial function in units of the reduced Planck mass $M_{\rm Pl}$ as $\tilde \phi = \phi/M_{\rm Pl}$ and the radial coordinate as $x = \mu r$. Setting $\tilde \omega = \omega/\mu$, this choice gives
\begin{eqnarray}
	\frac{1}{x}\frac{\mathrm{d}u}{\mathrm{d} x} &=& \frac{1 - e^u}{x^2} + \left(e^{-v}\tilde\omega^2 + 1\right) e^u \tilde\phi^2 + \left(\frac{\mathrm{d}\tilde\phi}{\mathrm{d}x}\right)^2\,,\label{eq:u}\\
	\frac{1}{x}\frac{\mathrm{d}v}{\mathrm{d} x} &=& \frac{e^u - 1}{x^2} + \left(e^{-v}\tilde\omega^2 - 1\right) e^u \tilde\phi^2 + \left(\frac{\mathrm{d}\tilde\phi}{\mathrm{d}x}\right)^2\,,\label{eq:v}\\
	\frac{\mathrm{d}^2\tilde\phi}{\mathrm{d}x^2} &=& -\left(1 + e^u - x^2 e^u \tilde\phi^2\right)\frac{1}{x} \frac{\mathrm{d}\tilde\phi}{\mathrm{d}x} + e^u (1 - e^{-v}\tilde\omega^2) \tilde\phi\,.\label{eq:phi}
\end{eqnarray}
The boundary conditions implemented to solve the system of equations above is
\begin{eqnarray}
	\lim_{r\to 0}\phi(r) &=& \phi_c\,,\qquad \lim_{r\to +\infty} \phi(r) = 0\,,\\
		\lim_{r\to 0}v(r) &=& v_c\,,\qquad \lim_{r\to +\infty}v(r) = 0\,,\\
	\lim_{r\to 0} \mathcal{M}(r) &=& 0\,,\qquad \lim_{r\to +\infty} \mathcal{M}(r) = M\,,
\end{eqnarray}
where $v_c$, $\phi_c$ are constants and where $\mathcal{M}(r)$ is defined in Eq.~\eqref{eq:enclmass}. These boundary conditions assure that the solution is localized in space. The system of equations with the boundary conditions can be solved numerically by integrating from the origin outward the set of Eqs.~\eqref{eq:u}-\eqref{eq:phi} and using a shooting method to satisfy the boundary conditions at infinity, to find stationary gravitational solutions~\cite{Dias:2015nua}. Given the amplitude of the scalar field at the core, $\phi(r = 0) \equiv \phi_c$, these conditions are satisfied for a set of $n$ frequency eigenvalues, where the $n$-th mode has $n-1$ nodes. We solve for the ground state $n = 1$, while higher values of $n$ correspond to the excited radial modes of the BS. 

Starting from a small value of $\phi_c \ll 1$, an increase in $\phi_c$ corresponds to an increase in the mass of the BS and to a decrease in its radius. Such a trend continues as long as the mass lies below a critical value~\cite{Breit:1983nr}
\begin{equation}
	\label{eq:kauplimit}
	M_{\rm Kaup} \approx 0.633 \,m_{\rm Pl}^2/\mu\,,
\end{equation}
above which both the radius and the mass of the BS decrease when increasing $\phi_c$. This behavior is also found in fermion stars: for example, a neutron star (NS) possesses a maximal mass~\cite{Chamel:2013efa}, as confirmed from observations~\cite{Ozel:2012ax, Rezzolla:2017aly}.

Once the solution for $\phi(r)$ is provided, we define the properties of the BS, namely the Noether charge $N$ in Eq.~\eqref{eq:noethercharge}, the mass $M$, and the radius $R$ containing 99\% of the mass of the BS, as
\begin{eqnarray}
	N &=& 4\pi \,\int_0^{+\infty} r^2\,\omega\,e^{\frac{v-u}{2}}\,\phi^2(r)\,\mathrm{d}r\,,\\
	M &=& 4\pi \int_0^{+\infty} r^2\, \left[\left(e^{-v}\,\omega^2 + \mu^2\right)\,\phi^2 + e^{-u}(\phi')^2\right]\mathrm{d}r\,,\\
	0.99 M &=& 4\pi \int_0^R r^2\, \left[\left(e^{-v}\,\omega^2 + \mu^2\right)\,\phi^2 + e^{-u}(\phi')^2\right]\mathrm{d}r\,.
\end{eqnarray}
Note, that $R$ is defined implicitly in the latter expression. The definition of the BS radius is rather ambiguous since BSs have no surface and the distribution extends to infinity, leading to different definitions in the literature. The compactness of a star of mass $M$ and radius $R$ is $\mathrm{C} \equiv GM/R$. The compactness corresponding to the Kaup limit is $\mathrm{C} \simeq 0.08$~\cite{Kaup:1968zz}.

\begin{figure}
	\includegraphics[width = \linewidth]{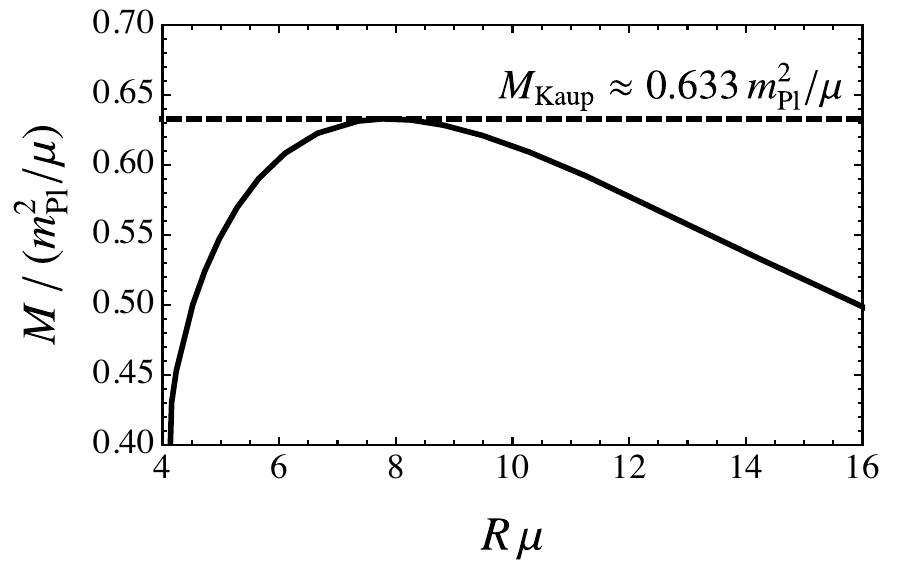} 
	\caption{{\it Solid black line}: Mass of the mini-boson star $M$ in units of $m_{\rm Pl}^2/\mu$, as a function of the radius $R$ in units of the inverse boson mass $\mu^{-1}$. {\it Dashed black line}: The mass of the mini-boson star corresponding to the Kaup limit, $M_{\rm Kaup} \approx 0.633\,m_{\rm Pl}^2/\mu$.}
	\label{fig:miniBS}
\end{figure}

\begin{figure}
	\includegraphics[width = 0.9\linewidth]{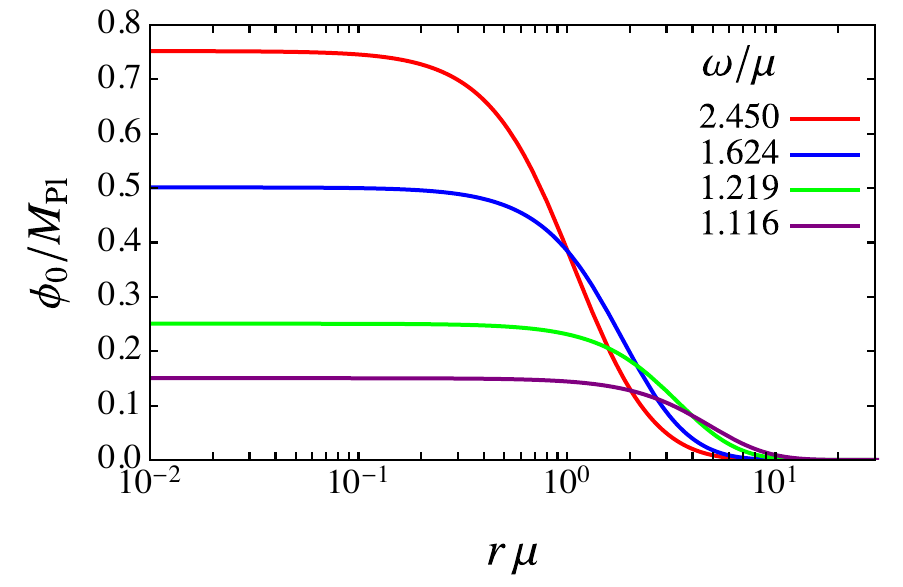} 
	\caption{The value of the boson field inside the BS in unit of $M_{\rm Pl}$, as a function of the radial coordinate in units of $\mu^{-1}$. Different values of the fundamental frequency $\omega$ are shown based on the color code in the figure.}
	\label{fig:PlotPhiBS}
\end{figure}

Fig.~\ref{fig:miniBS} shows the mass-radius relation obtained from the expressions above. The dashed line marks the Kaup mass in Eq.~\eqref{eq:kauplimit}. The curves are parametrized by the value of the field at the core $\phi_c$ as $M = M(\phi_c)$, $R = R(\phi_c)$, so that denser stars are more compact. This is shown in Fig.~\ref{fig:PlotPhiBS} sketching the radial profile as a function of the radial coordinate for a few representative values of $\tilde\omega=\omega/\mu$.

\subsection{Massive boson stars}
\label{sec:selfinteracting}

It is possible to alter the expression for the scalar field potential in Eq.~\eqref{eq:quadraticV} to produce a family of more massive BSs. A realistic extension of the BS potential might include higher-order nonlinear terms which result from bosonic self-interactions~\cite{Mielke:1980sa}. Here, we discuss the inclusion of a quartic term to the mini-boson star potential, as~\cite{Colpi:1986ye}
\begin{equation}
	\label{eq:selfinteracting}
	V(\Phi) = \mu^2|\Phi|^2 + \frac{\lambda}{2}\,|\Phi|^4\,,
\end{equation}
where $\lambda$ is a dimensionless coupling constant so that the self-interaction is repulsive for $\lambda \geq 0$. The structure of the resulting BS differs greatly from the case of the mini-boson star even for $|\lambda| \ll 1$. More precisely, the coupling constant can only be ignored for $|\lambda| \ll (\mu/m_{\rm Pl})^2$, which is not the case for a ``natural'' value $|\lambda| \sim 1$ and for a boson mass $\mu \ll m_{\rm Pl}$. Taking an effective field theory with the potential in Eq.~\eqref{eq:selfinteracting}, we expect to be in the ``strong'' regime in which the quartic coupling plays an important role.

With the potential in Eq.~\eqref{eq:selfinteracting}, the analogous of Eqs.~\eqref{eq:u}-\eqref{eq:phi} in the rescaled variables $x \equiv r\mu$ and $\tilde \phi = \phi/M_{\rm Pl}$ read
\begin{eqnarray}
	\frac{1}{x}\frac{\mathrm{d}u}{\mathrm{d} x} &=& \frac{1 - e^u}{x^2} + \left(e^{-v}\tilde\omega^2 + f(\tilde\phi)\right) e^u \tilde\phi^2 + \left(\frac{\mathrm{d}\tilde\phi}{\mathrm{d}x}\right)^2\,,\label{eq:u1}\\
	\frac{1}{x}\frac{\mathrm{d}v}{\mathrm{d} x} &=& \frac{e^u - 1}{x^2} + \left(e^{-v}\tilde\omega^2 - f(\tilde\phi)\right) e^u \tilde\phi^2 + \left(\frac{\mathrm{d}\tilde\phi}{\mathrm{d}x}\right)^2\,,\label{eq:v1}\\
	\frac{\mathrm{d}^2\tilde\phi}{\mathrm{d}x^2} &=& -\left(1 + e^u - x^2 e^u\,f(\tilde\phi)\, \tilde\phi^2\right)\frac{1}{x} \frac{\mathrm{d}\tilde\phi}{\mathrm{d}x} + e^u (2f(\tilde\phi)-1 - e^{-v}\tilde\omega^2) \tilde\phi\,,\label{eq:phi1}
\end{eqnarray}
where $f(\tilde\phi) = 1+\lambda'\tilde\phi^2/2$ and $\lambda' = \lambda/(8\pi G\mu^2)$. In the limit of strong coupling $|\lambda'| \gg 1$, we find the maximal mass for a spherically symmetric solution
\begin{equation}
	\label{eq:colpilimit}
	M_{\rm max} \approx 0.062 \sqrt{\lambda} \,m_{\rm Pl}^3/\mu^2\,.
\end{equation}
Note, that in Ref.~[\citen{Colpi:1986ye}] the relation is expressed as $M_{\rm max} \approx 0.22\Lambda^{1/2}\,m_{\rm Pl}^2/\mu$ with $\Lambda \equiv \lambda/(4\pi G\mu^2) = 2\lambda'$. In Table~\ref{tabParams} below, this expression is quoted as $M_{\rm max} \approx 0.31 \sqrt{\lambda'} \,m_{\rm Pl}^2/\mu$. The maximal mass for BSs in this model is generally much heavier than the Kaup limit in Eq.~\eqref{eq:kauplimit} for mini-boson stars, due to the different dependence on $\mu$ in Eq.~\eqref{eq:colpilimit}. The compactness of these massive BSs can reach that of a NS, $\mathrm{C}_{\rm max} \simeq 0.158$~\cite{2009PhRvD..80h4023G, Amaro-Seoane:2010pks, Chavanis:2011cz}. A fit of the size of galaxies and galaxy clusters might constrain the parameter space of BSs~\cite{Pires:2012yr, Souza:2014sgy, Chavanis:2020rdo}.

\begin{figure}
	\includegraphics[width = 0.9\linewidth]{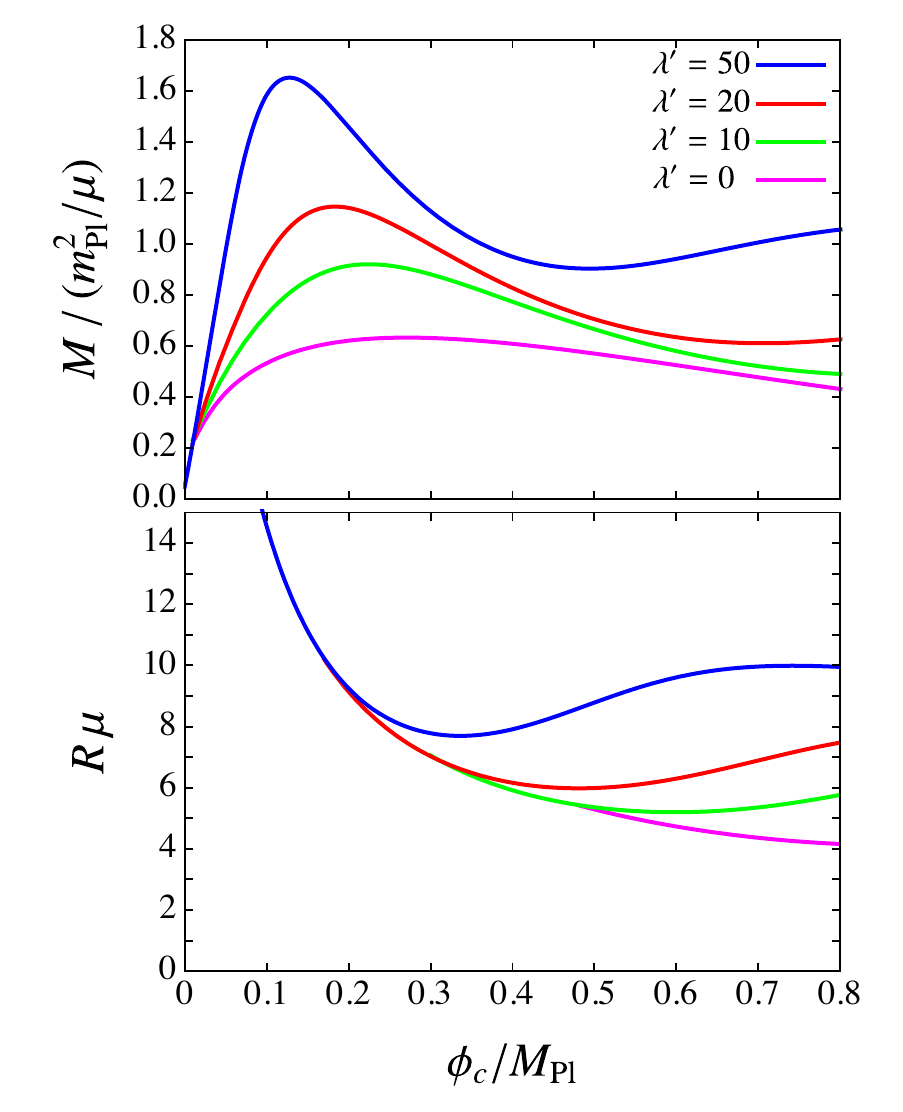} 
	\caption{Mass ({\it top panel}) and radius ({\it bottom panel}) of a BS described by the potential in Eq.~\eqref{eq:selfinteracting} as a function of the core value of the field, for the values of the coupling $\lambda' = 0$ (magenta), $\lambda' = 10$ (green), $\lambda' = 20$ (red), $\lambda' = 50$ (blue).}
	\label{fig:massiveBS}
\end{figure}

Fig.~\ref{fig:massiveBS} shows the mass in unit of $m_{\rm Pl}^2/\mu$ ({\it top panel}) and the radius in units of $\mu^{-1}$ ({\it bottom panel}) of a BS described by the potential in Eq.~\eqref{eq:selfinteracting}, as a function of the core value of the field $\phi_c$ in units of the reduced Planck mass $M_{\rm Pl} = m_{\rm Pl}/\sqrt{8\pi}$. The coloring codes the different values of the coupling used, with $\lambda' = 0$ (magenta), $\lambda' = 10$ (green), $\lambda' = 20$ (red), $\lambda' = 50$ (blue). The magenta line corresponds to the mini-boson star discussed in Sec.~\ref{sec:miniBS}. The mass shows a maximal value corresponding to the result in Eq.~\eqref{eq:colpilimit}.
Along with the quartic self-interacting potential, massive and stable stars can also be obtained by considering two massive fields $\Phi_1$ and $\Phi_2$, each described by the Lagrangian in Eq.~\eqref{eq:action} and with an additional non-gravitational interactions between the two scalars $\Phi_1^2\Phi_2^2$ which leads to peculiar phenomenology which can be tested in GW detectors~\cite{Guo:2020tla}.

\subsection{Non-gravitational solitons}
\label{sec:nongravitationalBS}

In theories of a complex scalar field with self-interactions, stationary non-gravitational soliton solutions named Q-balls can form in which the self-interacting potential replaces the gravitational interaction. Here, Q stands for the conserved charge associated to the conserved current within the U(1) theory. A Q-ball is a localized object existing in a flat spacetime, as opposed to a BS which is supported by its strong self-gravity~\cite{Tamaki:2010zz, Tamaki:2011zza}. A specific form of the self-interacting potential that allows for the existence of non-topological localized solitons in the absence of gravity is~\cite{Friedberg:1986tp, Friedberg:1986tq, Lee:1991ax}
\begin{equation}
	\label{eq:solitonic}
	V(\Phi) = \mu^2|\Phi|^2\,\left(1 - \frac{2|\Phi|^2}{\sigma_0^2}\right)^2\,,
\end{equation}
and deviates from Eq.~\eqref{eq:selfinteracting} in that the scalar field appears up to its sixth power. The constant $\sigma_0$ parametrizes the false vacuum solution at $|\Phi_0| = \sigma_0/\sqrt{2}$, with the true vacuum $|\Phi| \sim 0$ outside of the soliton being separate from the false vacuum inside $|\Phi_0|$ inside the star by a potential surface of width $\sim \mu^{-1}$. The potential in Eq.~\eqref{eq:solitonic} produces extremely compact stars with a maximal compactness $\mathrm{C}_{\rm max} \simeq 0.349$ which, being greater than $1/3$, allows for the existence of a photon sphere around the star. In the limit of strong coupling $\sigma_0 \ll m_{\rm Pl}$, the maximal mass for the soliton star is
\begin{equation}
	\label{eq:leelimit}
	M_{\rm max} \approx 0.0198 \sqrt{\lambda} \,m_{\rm Pl}^4/(\mu\sigma_0^2)\,.
\end{equation}
Note, that it is possible for a boson star to possess a light ring even for a smaller compactness $\mathrm{C} < 1/3$, as shown in explicit solutions of (unstable) mini-boson stars~\cite{Cunha:2017wao, Urbano:2018nrs} and oscillons~\cite{Macedo:2013jja}. It has been generally shown that any horizonless compact objects whose energy-momentum tensor satisfies $T^{\alpha\beta}p_\alpha p_\beta$ for any null vector $p_\alpha$, and that possesses an unstable light ring also has a stable light ring~\cite{Cunha:2017qtt}.

A different concept of Q-ball leading to the formation of an soliton is based on the idea~\cite{Coleman:1985ki}, where Q stands for the conserved charge associated to the conserved current within a theory with two real scalar fields enjoying a SO(2) symmetry. In this view, the solitons described in Eq.~\eqref{eq:ansatz} correspond to radial oscillations in the complex $\Phi$ plane, while these Q-balls correspond to circular motion in the complex $\Phi$ plane. For a complex scalar field $\Phi$ with a global U(1) symmetry, a Q-ball can form if the potential $V(|\Phi|^2)/|\Phi|^2$ shows a global minimum at $|\Phi| \neq 0$~\cite{Coleman:1985ki}. One such example is~\cite{Volkov:2002aj}
\begin{equation}
	\label{eq:solitonic1}
	V(\Phi) = \mu^2|\Phi|^2 + \lambda\left(|\Phi|^6 - a|\Phi|^4\right)\,,
\end{equation}
where $\lambda$ and $a$ are constants. Any potential that could lead to the formation of Q-balls should satisfy the condition $V''(0) > 2V(\Phi)/|\Phi|^2$~\cite{Coleman:1985ki}, which is not fulfilled by Eq.~\eqref{eq:selfinteracting}, but it is met by the potential in Eq.~\eqref{eq:solitonic1} for $\lambda > 0$.

A similar concept has been discussed in the theory describing an ``abnormal state'' of nuclear matter~\cite{Lee:1974ma,Lee:1974kn,Lee:1977xg} which could appear for sufficiently high nuclear densities, such that a new $\sigma$ field coupling to nucleons provide a negative contribution to the nucleon bare mass, leading to nearly massless nucleons. Since an unbroken continuous symmetry exists for $\sigma$ along with the associated conservation law, an alternative non-topological configuration called a Q-star can also be formed~\cite{Lynn:1988rb, Lynn:1989xb, Bahcall:1989yj, Hochron:1992uf}. Supersymmetric SM extensions predict the existence of baryonic Q-balls, macroscopic self-bound clumps of scalar particles which possess a defined electric and baryonic charge and which have been proposed as the dark matter~\cite{Kusenko:1997si}.

\section{Real scalar fields}
\label{sec:real}

\subsection{Oscillatons}

Unlike a complex scalar field, a real scalar field does not possess a U(1) symmetry granting the existence of a conserved charged. Therefore, stability would be lacking for this case. Nevertheless, soliton solutions for the field equation involving scalar bosons exist even without an explicit conserved Noether current~\cite{Seidel:1991zh, Copeland:1995fq}. These are known in the literature as \textit{oscillatons}. In contrast to the assumptions for a static metric adopted so far, oscillaton solutions are possible if the spacetime metric is time dependent and oscillatory, similarly to the breather solution of the sine-Gordon equation~\cite{Ablowitz:1973fn}. Oscillatons are then time dependent, non-topological, non-singular, asymptotically flat solutions to EKG~\cite{Urena-Lopez:2001zjo}. As a consequence, solitonic solutions are metastable, although their decay timescale can vastly exceed the age of the Universe.\footnote{\,See Ref.~[\citen{Urena-Lopez:2012udq}] for a similar study with a quartic potential.}

A minimally coupled real scalar field $\phi$ evolving on the background metric $g_{\alpha\beta}$ is described by the action
\begin{equation}
	\label{eq:actionB}
	\mathcal{S} = \int \mathrm{d}^4x\sqrt{-g} \left[\frac{\mathcal{R}}{16\pi G} - \frac{1}{2}\nabla^\alpha\phi\nabla_\alpha\phi - V(\phi)\right]\,,
\end{equation}
where $V(\phi)$ is the potential for the real scalar field. From the above action follows the Klein-Gordon equation,
\begin{equation}
	\label{eq:KGEB}
	\nabla^\alpha\nabla_\alpha\phi = \frac{\mathrm{d}V(\phi)}{\mathrm{d}\phi}\,.
\end{equation}
The variation of the action in Eq.~\eqref{eq:actionB} with respect to the metric tensor leads to the Einstein equation in Eq.~\eqref{eq:Einstein}, where the energy-momentum tensor $T_{\alpha\beta}^\Phi$ for the complex scalar field is replaced by
\begin{equation}
	\label{eq:energymomentumB}
	T_{\alpha\beta}^\phi = \nabla_\alpha\phi \nabla_\beta\phi - g_{\alpha\beta}\left(\frac{1}{2}\nabla^\gamma\phi\nabla_\gamma\phi + V(\phi)\right)\,.
\end{equation}

For a spherically-symmetric solution, we assume the metric parametrization in Eq.~\eqref{eq:metric} with the two time-dependent functions $u = u(r, t)$ and $v=v(r, t)$. Computing the Einstein tensor of the metric and equating it to the right-hand side of Eq.~\eqref{eq:Einstein} gives
\begin{eqnarray}
	e^{-u}\left(\frac{u'}{r} - \frac{1}{r^2}\right) + \frac{1}{r^2} &=& 8\pi G\,\left(\frac{1}{2}e^{-v}\,\dot\phi^2 + \frac{1}{2}e^{-u}(\phi')^2 + V(\phi)\right)\,,\\
	e^{-u}\left(\frac{v'}{r} + \frac{1}{r^2}\right) - \frac{1}{r^2} &=& 8\pi G \,\left(\frac{1}{2}e^{-v}\,\dot\phi^2 + \frac{1}{2}e^{-u}(\phi')^2 - V(\phi)\right)\,,\\
	\phi''(r) + \left(\frac{2}{r} + \frac{v'-u'}{2}\right)\phi'(r) &=& e^{u}\left[\frac{\mathrm{d}V(\phi)}{\mathrm{d}\phi} + e^{-v}\left(\ddot\phi + \frac{\dot u - \dot v}{2}\dot\phi\right)\right]\,.\label{eq:motion_tdep}
\end{eqnarray}
where a prime is a differentiation with respect to $r$ and a dot is a differentiation with respect to $t$. This set of equations differs from the corresponding Eqs.~\eqref{eq:u}-\eqref{eq:phi} for a complex scalar field because of the appearance of the time derivatives of the metric terms $\dot u$, $\dot v$. Oscillatons show a critical mass $M_c = 0.605\,m_{\rm Pl}^2/\mu$ and compactness $\mathrm{C} = 0.14$~\cite{Alcubierre:2003sx}, that distinguishes between the stable and unstable configurations as discussed in Sec.~\ref{sec:stability}.

\subsection{Axion stars}
\label{sec:axionstars}

The axion~\cite{Weinberg:1977ma, Wilczek:1977pj} is the pseudo-Goldstone boson associated with the spontaneous breaking of a global U(1) symmetry, first introduced by Peccei and Quinn (PQ)~\cite{Peccei:1977ur, Peccei:1977hh}. The complex PQ field after the spontaneous symmetry breaking reads $\Phi = f_a/\sqrt{2}\,\exp(i\phi/f_a)$, where $\phi$ is the axion field and $f_a$ is a new energy scale, the axion decay constant. The axion might play a crucial role in the evolution of the Universe as it could explain the dark matter observed~\cite{Preskill:1982cy, Abbott:1982af, Dine:1982ah, Stecker:1982ws}. See Refs.~[\citen{Marsh:2015xka, DiLuzio:2020wdo}] for recent reviews. For this reason, understanding the formation and evolution of compact objects formed within the QCD axion theory is crucial, as they would come with unique astrophysics signatures that can be searched to claim detectability, jointly with other axion-induced effects in the laboratory.

Axion stars are localized structures made of QCD axions and solutions to EKG. Despite axion stars being generally self-gravitating objects~\cite{Barranco:2010ib}, an important role in their equilibrium is played by self-interactions. The effects of non-homogeneities in the primordial axion field have also been discussed~\cite{Sakharov:1994id, Sakharov:1996xg, Khlopov:1998uj}. Axion stars should not be confused with compact clumps made of virialized axions called miniclusters~\cite{Hogan:1988mp, Kolb:1993zz, Kolb:1993hw}. Recent developments are in Refs.~[\citen{Visinelli:2018wza, Vaquero:2018tib, Ellis:2020gtq}].

The evolution of the axion field $\phi$ is described by the Klein-Gordon Eq.~\eqref{eq:KGEB} under a periodic potential which can be approximated when $\phi\ll f_a$ as
\begin{equation}
	V(\phi) = \frac{1}{2}\mu^2\phi^2 - \frac{\lambda}{24}\phi^4\,.
\end{equation}
Here, the mass of the axion is $\mu = \Lambda_a^2/f_a$ with $\Lambda_a \simeq 75.5\,$MeV, and the dimensionless coupling for the attractive self-interaction is $\lambda = (g_4 \mu/f_a)^2$ with $g_4 \simeq 0.59$~\cite{GrillidiCortona:2015jxo}. We expand the axion field in terms of a non-relativistic wave function
\begin{equation}
	\phi({\bf r}, t) = \frac{f_a}{\sqrt{2}}\left[\psi({\bf r}, t)e^{-i\mu t}+ \psi^*({\bf r}, t)e^{i\mu t}\right]\,,
\end{equation}
where the dimensionless function $\psi = \psi({\bf r}, t)$ varies slowly with time. With this representation for $\phi$, the Klein-Gordon Eq.~\eqref{eq:KGEB} reduces to the Schr{\"o}dinger equation,
\begin{equation}
	\label{eq:schroedinger}
	i\frac{\partial \psi}{\partial t} = -\frac{1}{2\mu}\nabla^2\psi + \mu\left(V_{\rm grav} + V_{\rm self}\right)\psi\,,
\end{equation}
describing the motion of the axion under the influence of the self-interaction potential $V_{\rm self}= - (g_4^2/8)|\psi|^2$. The gravitational potential $V_{\rm grav}$ is given by the Poisson equation
\begin{equation}
	\label{eq:poisson}
	\nabla^2 V_{\rm grav} = 4\pi G \rho\,,
\end{equation}
where the energy density of the non-relativistic axion is $\rho = \Lambda_a^4 |\psi|^2$. The set of Eqs.~\eqref{eq:schroedinger}-\eqref{eq:poisson} forms the Schr{\"o}dinger-Poisson system and it is analogous to the non-relativistic limit of the Gross-Pitaevskii equation when self-interactions are included~\cite{Chavanis:2011zi, 2011PhRvD..84d3532C}.

Due to the non-relativistic assumptions, the axion star behaves like an isotropic fluid for which the mass-radius relation in Eq.~\eqref{eq:massradiusrelation} holds. Axions stars for which self-gravity is counteracted by the effects of quantum pressure are in the ``dilute'' phase and are stable. Such an equilibrium is spoiled by self-interactions, which destabilize the star when the two potential terms in Eq.~\eqref{eq:schroedinger} match. This occurs at a critical mass~\cite{Chavanis:2011zi, Eby:2014fya, Visinelli:2017ooc}
\begin{equation}
	\label{eq:maxaxionstarmass}
	M_{\rm max} \approx \frac{6.6\times 10^{-10}\, M_\odot}{g_4} \, \left(\frac{\rm \mu eV}{\mu}\right)^2\,,
\end{equation}
above which no axion star solution is possible. This result is smaller than the maximal oscillaton mass for $\mu \gtrsim 10^{-11}\,$eV.

In the literature, compact solutions known as ``dense'' axion stars have been discussed within the Gross-Pitaevskii formalism~\cite{Braaten:2015eeu}. However, these solutions do not behave consistently due to large relativistic corrections and number-changing processes~\cite{Visinelli:2017ooc, Chavanis:2017loo, Schiappacasse:2017ham, Eby:2017teq, Eby:2019ntd}. Because of the importance of the QCD axion in cosmology and astrophysics, the formation and the properties of axion stars are being intensively studied. See Sec.~\ref{sec:formation} for the formation of BSs and Sec.~\ref{sec:othersearches} for detection signatures.

\section{Rotating boson stars}
\label{sec:rotatingBS}

So far, we have restricted the discussion to spherically symmetric configurations. We now consider the possibility that a BS possesses a spin and rotates around a fixed axis in space. The metric decomposition for a rotating object in GR, such as a Kerr BH or a spinning BS, can be generally written as
\begin{equation}
	\label{eq:metric_rotating0}
	\mathrm{d}s^2 = -N_t^2\, \mathrm{d}t^2 + \gamma_{ij}\left(\mathrm{d}x^i +\beta^i\mathrm{d}t\right)\left(\mathrm{d}x^j +\beta^j\mathrm{d}t\right)\,,
\end{equation}
where $\gamma_{ij}$ is the metric induced on a hypersurface $\Sigma_t$ of constant $t$ by the metric $g_{\alpha\beta}$, $\beta^i$ is the shift vector, and $N_t$ is the lapse function so that the orthogonal direction to $\Sigma_t$ is $(\partial/\partial t)^i = N_t n^i+\beta^i$, with $\gamma_{ij}n^i\beta^j = 0$. This line element contains a larger number of functions with respect to the decomposition in Eq.~\eqref{eq:metric}, because the non-zero rotation reduces the symmetry of the problem. For a stationary and axisymmetric spacetime, where the surfaces of constant $(r, \vartheta)$ are orthogonal to the surfaces of constant $(t, \varphi)$, the metric in Eq.~\eqref{eq:metric_rotating0} reduces to
\begin{equation}
	\label{eq:metric_rotating}
	\mathrm{d}s^2 = -e^{v}\, \mathrm{d}t^2 + e^{u}\, \left(\mathrm{d}r^2 + r^2\mathrm{d}\vartheta^2\right) + e^{w}\, r^2\sin^2\vartheta\left(\mathrm{d}\varphi + \beta\mathrm{d}t\right)^2\,,
\end{equation}
where the shift vector is $\beta^i = (0,0,\beta)$ and where $u$, $v$, $w$ are functions of $r$ and $\vartheta$ only.

The angular momentum of a BS is quantized in units of an ``azimuthal'' quantum number $m$ which enters the decomposition of the wave function as~\cite{Silveira:1995dh, Schunck:1996he, Yoshida:1997nd, Yoshida:1997qf}
\begin{equation}
	\label{eq:ansatz_rotating}
	\Phi = \phi(r, \vartheta)\,e^{-i\omega t}\, e^{-i m \varphi}\,.
\end{equation}
Solutions for rotating BSs are found from the EKG and depend on the choice of the boson potential. Rotating solutions in a relativistic setup have been discussed for mini-boson stars with $m =1$ to 10 and for $m = 500$ in the weakly relativistic regime~\cite{Schunck:1996he, Mielke:1997re}, for $m =1$, $m =2$ in the strongly relativistic regime~\cite{Yoshida:1997nd, Yoshida:1997qf}, and in the presence of a large self-interaction for static axisymmetric configurations for $m \gg 1$~\cite{Ryan:1996nk}. Rotating BS solutions with the solitonic potential in Eq.~\eqref{eq:solitonic1} have been computed for $m = 1$~\cite{Volkov:2002aj, Kleihaus:2005me}. Following the results for non-rotating BSs~\cite{Guerra:2019srj}, a rotating BS with a periodic potential has also been studied~\cite{Delgado:2020udb}. Rotating axion stars could increase their mass by a factor $\mathcal{O}(10)$ compared to the results in Sec.~\ref{sec:axionstars} and could be a major contribution to the dark matter (DM)~\cite{Davidson:2016uok}.

Solutions for a rotating BS possess a higher maximal mass than the non-rotating counterparts, for example the Kaup limit for the mini-boson star, see Eq.~\eqref{eq:kauplimit}, and the maximal mass for a massive BS in Eq.~\eqref{eq:colpilimit} modify as in Table~\ref{tabParams} for different values of $m$. As $m$ increases, the star transitions from a slowly rotating configuration to a highly relativistic BS. This affects the motion of test particles through the Lense-Thirring effect, which can serve as an indication for the existence of these objects~\cite{Zhang:2021xhp}.
\begin{table}[tb!]
\def\arraystretch{1.3}
	\centering
	\tbl{The maximal mass of the boson star that can be achieved for different values of $m$.}{
	\begin{tabular}{|cccl|cccl|}
    \cline{2-4}
    \hline\hline
		Model & $m$ & $M_{\rm Kaup}\,[m_{\rm Pl}^2/\mu]$ & Reference & Model & $m$ & $M_{\rm max}\,[m_{\rm Pl}^2/\mu]$ & Reference \\
		 \hline
		Mini-boson star &0 & 0.6630 & [\citen{Kaup:1968zz, Breit:1983nr}] & Massive BS &0 & $0.31 \sqrt{\lambda'}$ & [\citen{Colpi:1986ye}]\\ 
		&1 & 1.3155 &[\citen{Yoshida:1997qf, Grandclement:2014msa}] & ($\lambda' = 100$) &1 & 3.14 &[\citen{Grandclement:2014msa}]\\
		&2 & 2.2159 &[\citen{Grandclement:2014msa}] & &2 & 3.48 &[\citen{Grandclement:2014msa}]\\
		&3 & 3.5287 &[\citen{Grandclement:2014msa, Ontanon:2021hbg}] & &3 & 4.08 &[\citen{Grandclement:2014msa}]\\
		&4 & 5.0590 &[\citen{Grandclement:2014msa, Ontanon:2021hbg}] & &4 & 5.59 &[\citen{Grandclement:2014msa}]\\
		&5 & 6.6681 &[\citen{Grandclement:2014msa, Ontanon:2021hbg}] & & & &\\
		&6 & 8.2824 &[\citen{Grandclement:2014msa, Ontanon:2021hbg}] & & & &\\
	\hline\hline
	\end{tabular}}
	\label{tabParams}
\end{table}

In general, the slow rotation limit of a BS cannot be taken because of the quantization of angular momentum~\cite{Kobayashi:1994qi}. Solutions exists in the non-relativistic approximation, where rotation can be thought as a perturbation carrying the angular momentum over the background potential ground state~\cite{Ferrell:1989kz}. In this approach, the equations reduce to the Newtonian case and stable rotating configurations with axial symmetry exist~\cite{Silveira:1995dh}.

A rotating BS possesses an axisymmetric solution which is topologically distinct from the spherically symmetric solutions discussed for non-rotating BSs. In fact, the velocity field ${\bf v}$ is related to the wave function $\psi({\bf r}, t)$ as~\cite{LifshitzBook9} [see also Eq.~\eqref{eq:defv} below]
\begin{equation}
	{\bf v} = {\bf \nabla}\,{\rm arg} \,\psi({\bf r}, t)/\mu\,.
\end{equation}
If the density distribution is non-zero everywhere, the configuration is irrotational since ${\bf \nabla} \times {\bf v} = 0$, so that the energy density of a rotating BS must be concentrated within a torus~\cite{Ryan:1996nk, Schunck:1996he, Grandclement:2014msa, Dmitriev:2021utv}. This is seen from the explicit computation of the wave function, whose dependence near the origin is $\phi(r, \vartheta) \propto r^m$ which vanishes for $m > 0$.

The stability of rotating BSs is currently under scrutiny. It has been shown that for non-relativistic stars (see e.g.\ Sec.~\ref{sec:axionstars}) and for a negligible or attractive self-interaction, a gravitationally bound BS would be unstable for any angular momentum, while a repulsive self-interaction allows for a rotating BS solution for $m = 1$. This is ultimately caused by transitions that conserve the total spin and allow to jump in values of the angular momentum given by $m$~\cite{Dmitriev:2021utv}. Any excited state would relax towards the ground state by the emission of relativistic bosons and potentially gravitational waves. Stability is further discussed in Sec.~\ref{sec:stability}.

\section{Additional configurations}
\label{sec:speculative}

\subsection{Massive spin-one bosons}
\label{sec:procastar}

A massive spin-one field $A^\alpha$, of mass $\mu$, is described by the Proca equation. Contrary to the photon, which is massless and whose modes are longitudinal with respect to the direction of the propagation, a massive spin-one field also possesses a transverse mode. The action for this theory is
\begin{equation}
	\label{eq:action_proca}
	\mathcal{S} = \int \mathrm{d}^4x\sqrt{-g} \left[\frac{\mathcal{R}}{16\pi G} -\frac{1}{4}F^{\alpha\beta}\,F^*_{\alpha\beta} - \frac{1}{2}\mu^2A^\alpha A_\alpha^*\right]\,,
\end{equation}
where $F^{\alpha\beta} = \nabla^\alpha A^\beta - \nabla^\beta A^\alpha$ is the field strength and where an asterisk means complex conjugation. The Proca field $A^\alpha$ is invariant under a U(1) symmetry, so that a conserved charge exists. Self-gravitating ``Proca star'' solutions exist with a maximal mass~\cite{Garfinkle:2003jf, Brito:2015pxa, Brito:2015yga}
\begin{equation}
	M_{\rm max} = 1.058\,m_{\rm Pl}^2/\mu\,,
\end{equation}
which is slightly higher than the corresponding Kaup limit for mini-boson stars in Eq.~\eqref{eq:kauplimit}. The maximal mass separates stable solutions from the unstable ones, see Sec.~\ref{sec:stability}. A full non-linear numerical evolutions of Proca stars reveals that three outcomes are possible for the unstable branch, namely (i) migration to the stable branch, (ii) dispersion of the scalar field, or (iii) collapse to a Schwarzschild black hole~\cite{Sanchis-Gual:2017bhw}. Along with rotating Proca stars~\cite{Brito:2015pxa}, an extension to Proca Q-balls and charged Proca stars~\cite{SalazarLandea:2016bys} or exotic configurations with a negative cosmological constant~\cite{Duarte:2016lig} have also been discussed. Along with the potential depending on the quantity $Y \equiv \bar A_\mu A^\mu$~\cite{Brito:2015pxa, Brihaye:2017inn, SalazarLandea:2016bys, Minamitsuji:2018kof, Herdeiro:2020jzx, Cardoso:2021ehg, Zhang:2021xxa, Jain:2022kwq}, the quartic self-interaction of the form $\propto \bar A_\mu \bar A_\mu A^\nu A_\nu - Y^2$ has also been recently considered~\cite{Aoki:2022mdn}. It has been recently shown that Proca star solutions exist even in regions of the parameter space in which the effective field theory suffers from a number of issues and breaks down, when a more complete theory with additional fields is employed~\cite{Clough:2022ygm, Coates:2022qia, Aoki:2022woy} or when suitable strategies can be designed at the level of the infrared theory~\cite{Barausse:2022rvg}.

A different configuration is obtained in vector-tensor theories beyond GR characterized by the Lagrangian
\begin{equation}
	\mathcal{L} = \frac{\mathcal{R}}{16\pi G} -\frac{1}{4}F^{\alpha\beta}\,F_{\alpha\beta} + \frac{1}{4} V_\mu V_\nu G^{\mu\nu}\,,
\end{equation}
where $V^\mu$ is a vector field of field strength $F^{\mu\nu} = \partial^\mu V^\nu - \partial^\nu V^\mu$ which is non-minimally coupled with the Einstein tensor $G^{\mu\nu}$~\cite{Tasinato:2022vop}. The interior region of a compact solution shows an anisotropic stress tensor $T^\mu_\nu = {\rm diag}(-\rho, p_r, p_t, p_t)$ with components
\begin{eqnarray}
	\rho &=& \frac{2\gamma}{(1+2\gamma)r^2}\,,\\
	p_r &=& 0\,,\\
	p_t &=& \frac{\gamma}{2}\rho\,,
\end{eqnarray}
depending on a parameter $\gamma$. The compactness of these solutions range between
\begin{equation}
	0 \leq \mathrm{C} \equiv  \frac{\gamma}{1+2\gamma} \leq 1/2\,,
\end{equation}
so that this configuration can mimic a compact BH in the limit $\gamma \gg 1$.

\subsection{Boson-fermion stars}
\label{sec:BFS}

A boson-fermion star is a localized equilibrium configuration composed of both bosons and fermions~\cite{Henriques:1989ar}. In the simplest picture, the energy-momentum tensor is expressed as the sum of a bosonic component given in Eq.~\eqref{eq:energymomentum} and a fermionic component described in the fluid formalism by
\begin{equation}
	T^F_{\alpha\beta} = (\rho + p)u_\alpha u_\beta + g_{\alpha\beta}p\,,
\end{equation}
where $u^\alpha$ is the fermion four-vector and $\rho$, $p$ are the energy density and pressure of the fermion component. For a fermion field, the fluid formulation is possible since the pressure term is isotropic.

In the presence of a boson-fermion interaction, the real and imaginary components of the boson field follow different equations~\cite{deSousa:1995ye}. In this case, it is not possible to treat the function $\phi(r)$ as real, contrary to what has been discussed below Eq.~\eqref{eq:ansatz}. A star composed of a mix of a boson and a fermion species shows peculiar properties. Contrary to pure boson stars, a boson-fermion star possesses a slow rotating solution even in the relativistic prescription~\cite{deSousa:2000eq} and their excited states appear to be generally more stable than the purely bosonic counterparts~\cite{DiGiovanni:2020frc, DiGiovanni:2021vlu}.

\subsection{Pion stars}

In the presence of isospin or strangeness, the low-energy properties of strongly interacting matter might be drastically altered. A non-zero isospin chemical potential $\mu_I$ or a strangeness chemical potential $\mu_S$ might lead to the formation of a pion ($\pi c$) or a kaon ($Kc$) condensate, along with the normal phase~\cite{Son:2000xc, Kogut:2001id}. For example, a system of charged pions with zero baryon and strangeness chemical potential and with $\mu_I \neq 0$ could condense into a zero momentum state and act as a superfluid if $\mu_I \geq m_\pi$, where $m_\pi$ is the vacuum pion mass, with the phase transition occurring at $\mu_I = m_\pi$. The stability of charged $\pi^+$ (for $\mu_I>0$) in the $\pi c$ phase allows for the possibility of pion stars formed of condensed pions, in which electric neutrality is granted by matching the isospin number density with the electron number density~\cite{Carignano:2016lxe, Carignano:2016rvs}. Solving the Tolman-Oppenheimer-Volkoff equation allows to recover the internal equation of state of the star as well as a stellar mass $\sim \mathcal{O}(10\,M_\odot)$ and a stellar radius $\sim \mathcal{O}(10{\rm \,km})$. The mass-radius relation and the equation of state can also be obtained without imposing electric charge neutrality and considering both a gas of electrons or muons along with the condensate, leading to considerably larger and more massive stars using recent lattice results~\cite{Brandt:2018bwq, Andersen:2018nzq}.

\subsection{Oscillons forming after inflation}

In models for single-field inflation, the Universe is dominated by a massive real scalar field $\phi$, the {\it inflaton}. At the end of this stage, the inflaton transfers its energy to lighter particles through various possible {\it reheating} processes, so that the Universe transitions to the standard $\Lambda$CDM cosmology. In models in which such a reheating occurs through parametric resonance, it is possible for the Universe to become dominated by oscillons resulting from localized configurations of the inflaton~\cite{Amin:2010jq, Amin:2011hj, Amin:2014eta, Lozanov:2017hjm, Hiramatsu:2020obh}. Numerical simulations in 3D show that oscillons generate through fragmentation of the inflaton condensate~\cite{Amin:2010dc} and are generally long-lived and stable against collapse~\cite{Amin:2010jq}. Once the field begins to oscillate coherently about the minimum, oscillons form if field perturbations grow enough for self-interactions to become important and the potential is shallower than quadratic, $\phi^{2\gamma}$ with $\gamma < 1$ so that the production of oscillons is energetically favored once the inflaton field fragments~\cite{Amin:2010dc, Antusch:2017flz}. Gravitational effects might be important during preheating and could lead to oscillon collapse, as shown in numerical relativity simulations~\cite{Kou:2019bbc}.

Oscillons might emerge in models in which the potential shows a plateau, such as those generated within string or supergravity scenarios~\cite{Silverstein:2008sg, McAllister:2008hb, Kallosh:2010ug, Adshead:2010mc}. The potential behaves as $\phi^{2\gamma}$ with $\gamma < 1$ during the slow-roll regime, to then transition to a quadratic potential when the reheating process begins at a scale $\phi \ll m_{\rm Pl}$. The oscillon-dominated regime is effectively a matter-dominated stage during which higher modes of the primordial power spectrum are enhanced, possibly leading to the formation of primordial BHs~\cite{Khlopov:1985jw} and to the generation of a primordial GW spectrum~\cite{Zhou:2013tsa}.

\subsection{Modified gravity and boson stars}

In a scalar-tensor theory, gravity is mediated by both a scalar and a tensor field, which in the Jordan-Brans-Dicke (JBD) framework is expressed by the action~\cite{Brans:1961sx}
\begin{equation}
    \label{eq:JBDaction}
    \mathcal{S} = \int \mathrm{d}^4 x \sqrt{-g}\left(\frac{1}{16\pi G} \,\left[\Psi\mathcal{R} - \frac{\omega}{\Psi}g^{\alpha\beta}\partial_\alpha\Psi\partial_\beta\Psi\right] - \nabla^\alpha\bar\Phi\nabla_\alpha\Phi - V(|\Phi|^2) \right)\,,
\end{equation}
where $\Psi$ is a new scalar field that replaces the gravitational constant and $\omega$ is a new parameter in the theory so that GR is recovered for $\omega \to +\infty$. The rest of the notation is the same as in Eq.~\eqref{eq:action}. Typically, massive BSs forming in the JBD framework are slightly lighter than their GR counterparts for given values of the parameters $\mu$ and $\lambda$ in Eq.~\eqref{eq:selfinteracting}~\cite{Gunderson:1993cq}. For example, for $\omega = 6$ the Kaup limit in Eq.~\eqref{eq:kauplimit} modifies as $M_{\rm max} \approx 0.600m_{\rm Pl}^2/\mu$, with the difference reducing for larger $\omega > 6$. The scalar-tensor gravity generalization of the JBD framework in which the effective gravitational constant is a variable $\omega = \omega(\Phi)$ allows for BS solutions with a wider mass range that could evolve with time~\cite{Torres:1997np, Torres:1997cvs, Comer:1997ns}.

A different approach leading to a modification of GR is Palatini gravity, in which the Ricci scalar in Eq.~\eqref{eq:action} is replaced with a generic function of $\mathcal{R}$,
\begin{equation}
	\label{eq:fRaction}
	\mathcal{S} = \int \mathrm{d}^4x\sqrt{-g} \left[\frac{f(\mathcal{R})}{16\pi G} - \nabla^\alpha\bar\Phi\nabla_\alpha\Phi - V(|\Phi|^2)\right]\,.
\end{equation}
For the quadratic theory $f(\mathcal{R}) = \mathcal{R} + \xi \mathcal{R}^2$, where $\xi$ is a coupling, the BSs produced in Palatini gravity for $\xi>0$ are fairly similar to those found in GR previously discussed~\cite{Maso-Ferrando:2021ngp}. The properties of BS have also been discussed in Horndeski theories where a coupling between the kinetic scalar term and Einstein tensor exists~\cite{Brihaye:2016lin, Verbin:2017bdo, Barranco:2021auj} and in theories with Gauss-Bonnet couplings~\cite{Hartmann:2013tca, Brihaye:2013zha, Baibhav:2016fot}. In models of a scalar field non-minimally coupled to matter, the so-called chameleon field, compact solitonic solutions can be formed, although such configurations are unstable~\cite{Dzhunushaliev:2011ma, Folomeev:2013hoa}.

\section{Bose-Einstein condensate}
\label{sec:BEC}

\subsection{Equations for the bulk distribution}

In a Bose-Einstein condensate (BEC) forming below a very low critical temperature, bosons occupy the ground state level of the system with minimum momentum. Under certain conditions, a BEC could be realized even at the galactic scale~\cite{Baldeschi:1983mq} and could model the DM observed in galaxies, as discussed in Sec.~\ref{sec:DM}. The evolution of a self-gravitating BEC with the addition of a self-interaction is governed by the Schr{\"o}dinger-Poisson system of Eqs.~\eqref{eq:schroedinger}-\eqref{eq:poisson}. The wavelike behavior of a particle of mass $\mu$ and wave function $\psi$ connects with the macroscopic interpretation in terms of an ensemble of particles in classical physics thanks to the Madelung transformation~\cite{1927ZPhy...40..322M}
\begin{equation}
	\label{eq:madelung}
	\psi = \sqrt{\rho/\mu} \,e^{i\mathcal{S}}\,,
\end{equation}
where $\rho$ is the density of the parcel of fluid considered and $\mathcal{S}$ is the action of the particle. The transformation encodes the interpretation of $|\psi|^2$ in terms of the number density of particles, and connects to the continuity equation once we define the three-dimensional velocity~\cite{LifshitzBook9}
\begin{equation}
	\label{eq:defv}
	{\bf v} = {\bf \nabla} \mathcal{S} / \mu\,.
\end{equation}
Inserting Eq.~\eqref{eq:madelung} into the Schr{\"o}dinger Eq.~\eqref{eq:schroedinger} yields two equations for the real and the imaginary parts. The imaginary part is the continuity equation in hydrodynamics,
\begin{equation}
	\label{continuity}
	\frac{\partial \rho}{\partial t} + \left({\bf v} \cdot {\bf \nabla}\right) \,\rho = -\rho\,\left({\bf \nabla}\cdot {\bf v}\right)\,,
\end{equation}
while the real part is the Hamilton-Jacobi equation for the action $\mathcal{S}$ and includes the ``quantum pressure'' term $Q$,
\begin{eqnarray}
	\frac{\partial \mathcal{S}}{\partial t} &=& -\frac{1}{2\mu}(\nabla \mathcal{S})^2 - \mu\left(V_{\rm grav} +V_{\rm self} \right) - Q\,.\label{newton}\\
	Q &=& -\frac{1}{2\mu}\frac{\nabla^2\sqrt{\rho}}{\sqrt{\rho}}\,.
\end{eqnarray}
Taking the gradient of Eq.~\eqref{newton} and using Eq.~\eqref{eq:defv} leads to an extended Euler equation that includes the quantum pressure,
\begin{equation}
	\label{eq:euler}
	\frac{\partial {\bf v}}{\partial t} + \left({\bf v} \cdot {\bf \nabla}\right) \,{\bf v} = -\frac{1}{\rho}{\bf \nabla} p -{\bf \nabla} V_{\rm grav} - \frac{1}{\mu}{\bf \nabla}Q\,,
\end{equation}
where the pressure due to the self-interacting term is $p = \rho V_{\rm self} - W_{\rm self}$ and $W_{\rm self}$ is a primitive function of $V_{\rm self}$.

In the steady-state regime in which ${\bf v} = 0$ and the properties of the fluid are independent of time, the gradient of Eq.~\eqref{eq:euler} with the Poisson Eq.~\eqref{eq:poisson} gives a generalization to the Lane-Emden equation~\cite{lane_h_j_1870_1450030, 1907gask.book.....E} that includes the quantum pressure term~\cite{Chavanis:2011zi, Visinelli:2015wha, Kouvaris:2015rea, Sarkar:2017aje}
\begin{equation}
	\label{eq:laneemden}
	\nabla\left(\frac{1}{\rho}{\bf \nabla} p\right) + 4\pi\,G\,\rho - \frac{1}{2\mu^2}\nabla^2\left(\frac{\nabla^2\sqrt{\rho}}{\sqrt{\rho}}\right) = 0\,,
\end{equation}
where the three terms to the left-hand side correspond to the effect of the pressure due to the particle self-scattering, the attraction from the self-gravity of the fluid, and the repulsion due to quantum pressure, respectively. The sign of the self-scattering potential depends on the nature of the interaction. For a null self-interaction, the expression describes the balance between gravity and the quantum pressure, which is the equilibrium attained in an isotropic BS, giving the mass-radius relation in Eq.~\eqref{eq:massradiusrelation}~\cite{Ruffini:1969qy, Membrado:1989bqo}. In the different regime in which the quantum pressure can be neglected, a repulsive self-interaction can sustain the equilibrium against self-collapse~\cite{Boehmer:2007um, Hui:2016ltb}.

\subsection{Jeans instabilities}
\label{sec:jeans}

We now discuss the dynamical instabilities in the linear regime of the self-gravitating condensate. We decompose the density in terms of a mean background plus a perturbation, $\rho = \bar\rho + \rho_1$, and we consider the velocity ${\bf v}$ to be of the same order as the linear perturbations. The Euler-Poisson Eqs.~\eqref{eq:poisson},~\eqref{continuity},~\eqref{eq:euler} read
\begin{eqnarray}
	{\bf \nabla}^2V_{\rm grav} &=& 4\pi\,G\,\rho_1\,,\label{poisson1}\\
	\frac{\partial \rho_1}{\partial t} &=& -\bar\rho\,\left({\bf \nabla}\cdot {\bf v}\right)\,,\label{continuity1}\\
	\bar\rho\frac{\partial {\bf v}}{\partial t} &=& - gc_s^2{\bf \nabla} \rho_1 - \bar\rho{\bf \nabla} V_{\rm grav} + \frac{1}{4\mu^2}{\bf \nabla}\left(\nabla^2\rho_1\right)\,,\label{eq:euler1}
\end{eqnarray}
where the sound speed squared is $c_s^2 = \delta p/\delta \rho$ and $g$ encodes the sign of the self-interaction, with the repulsive potential being $g = +1$. Taking the time derivative of Eq.~\eqref{continuity1}, inserting Eq.~\eqref{eq:euler1} and using Eq.~\eqref{poisson1} for the Fourier-expanded perturbation $\rho_1 \propto \exp[i({\bf k} \cdot {\bf r} - \omega t)]$ gives the dispersion relation~\cite{Chavanis:2011zi, Chavanis:2020jkc}
\begin{equation}
	\label{soundspeed}
	\omega^2 = \frac{k^4}{4\mu^2} + gc_s^2k^2 - 4\pi\,G\,\bar\rho\,.
\end{equation}
Perturbations grow for modes $k > k_J$, where the Jeans wave number $k_J$ is the real and positive root of $\omega = 0$. When pressure is neglected, the Jeans wave number is~\cite{Khlopov:1985jw, Bianchi:1990mha, Silverman:2002qx, Hu:2000ke}
\begin{equation}
	k_J = \left(16\pi\,G\mu^2\,\bar\rho\right)^{1/4}\,,
\end{equation}
which sets the limit for the condensate collapse discussed previously. A repulsive self-interacting potential can also halts the collapse against gravity, so that when quantum pressure can be neglected we obtain~\cite{Boehmer:2007um}
\begin{equation}
	k_J = \sqrt{\frac{4\pi\,G\,\bar\rho}{c_s^2}}\,.
\end{equation}
On the contrary, an attractive self-interaction can take over the attraction of gravity and counteract the effects of quantum pressure. In this configuration, the Jeans length is~\cite{Chavanis:2011zi, Harko:2014vya}
\begin{equation}
	k_J = 2\mu|c_s|\,.
\end{equation}
This scale is relevant when considering axion dark matter, whose self-interaction is attractive, which is expected to form low mass axion stars but not a dark matter halos of galactic size~\cite{Guth:2014hsa, Chavanis:2015zua, Chavanis:2016dab}.

\section{Formation of boson stars}
\label{sec:formation}

The assessment of the existence of BSs and oscillatons in our Universe demands a rigorous study of possible formation mechanisms. Starting from a boson cloud, the formation of a compact object requires dissipating the excess kinetic energy through some cooling mechanism. The formation of bosonic compact objects might occur in environments with a large density of fundamental scalars. This can occur either at the core of dense boson clouds through relaxation from an incoherent initial condition or in the early Universe if the scalar is a substantial component of the dark matter and initial coherent field oscillations.\footnote{\,For pseudo-scalar fields, a field excursion of $\mathrm{O}(1)$ could lead to the formation of solitons and oscillons in the early Universe~\cite{Arvanitaki:2019rax}.}

A self-gravitating halo of a bosonic field could eject scalars which drag away the excess kinetic energy and condense to form a compact solitonic object. This ``gravitational cooling'' process proceeds efficiently for complex scalar fields forming BSs, real scalar fields forming oscillons~\cite{Seidel:1993zk}, and for vector fields forming Proca stars~\cite{DiGiovanni:2018bvo}. Within spherical symmetry where no gravitational radiation is emitted, gravitational cooling is the main dissipative mechanism to form compact objects through the emission of bosons and to accrete solitons already present in the halo. In the early Universe, the formation of a BS out of a fundamental scalar relies on the mechanism of Jeans instability, which is the clumping effect discussed in Sec.~\ref{sec:jeans}. Stationary solutions are generally stable below the Jeans length $\lambda_J$, while for larger scales $\lambda > \lambda_J$ clumping proceeds as long as the linear approximation holds. Numerical simulations of the process follow the dynamical formation of BSs starting from a virialized halos of bosons~\cite{Levkov:2018kau}, which fragments to form isolated solitons of the size comparable to the Jeans length~\cite{Guth:2014hsa}. See Ref.~\citen{2021EPJP..136..703C} for a treatment of Bose-Einstein condensation in terms of the kinetic theory of quantum particles.

The gravitational relaxation time $\tau_{\rm gr}$ at which the low energy levels populates forming the soliton is obtained from the kinetic equation for the energy distribution of bosons~\cite{Levkov:2018kau},
\begin{equation}
	\tau_{\rm gr} \sim \frac{\sqrt{2}}{12\pi^3}\,\frac{\mu v^6}{G^2n^2\ln p}\,,
\end{equation}
where $R$ is the size of the host halo with mean boson particle density $n$, $v$ is the velocity dispersion, and $p \equiv R\mu v$. The kinetic regime is characterized by $p \gg 1$ and corresponds to a large velocity dispersion within the halo. Setting the virial velocity $v^2 \sim 8 \pi G m n R^2 /3$ gives $\tau_{\rm gr} \propto (R/v) p^3/\ln p$ and the relaxation time is proportional to the free fall time. When $p \sim 1$, the kinetic regime cannot be applied and the soliton forms immediately, as seen in cosmological simulations with coherent initial conditions~\cite{Schive:2014dra, Schive:2014hza, Schive:2015kza, Schwabe:2016rze}. The inclusion of a self-interacting potential in the relaxation process shows that the gravitational cooling still proceeds through the emission of scalars and produces a spherically symmetric configuration even when the initial condition is non-spherical~\cite{Guzman:2004wj, Guzman:2006yc}. In solitons of galactic size, the core density undergoes quasi-coherent oscillations~\cite{Veltmaat:2018dfz} whose motion would affect stellar dynamics, heat up the central regions of galaxies, and halt the inspiral of very massive objects due to dynamical friction, leading to strong constraints on the mass of ultralight bosons~\cite{Marsh:2018zyw, 2019ApJ...871...28B, Marsh:2021lqg}.

Modifying the initial conditions to account for the highly incoherent initial conditions while using similar numerical techniques leads to similar results~\cite{Levkov:2018kau}. For example, this framework can describe the formation of miniclusters, dense virialized clumps formed of the axion field in the early Universe~\cite{Hogan:1988mp, Kolb:1993zz, Kolb:1993hw}. Inside the dense environment of an axion minicluster in an expanding cosmological background, the self-interaction of the axion potential leads to the formation of localized dense pseudo-soliton configurations known as axitons~\cite{Kolb:1993hw, Vaquero:2018tib}. Locally, the gravitational cooling mechanism of axions inside a minicluster leads to the formation of an axion star~\cite{Eggemeier:2019jsu} similarly to a solitonic core for light DM, while the outer profile follows a self-similar profile $\propto r^{-9/4}$~\cite{Bertschinger:1985pd}. Comparing the velocity of the axion from the Heisenberg uncertainty $v \sim (\mu R)^{-1}$ with the virial radius $R \sim G M/v^2$ gives the virial velocity of the axion star $v_{\rm vir} \sim GM\mu$~\cite{Hui:2016ltb}. The axion star growth saturates once $v_{\rm vir}$ equates the internal velocity of the axions in the minicluster (mc),
\begin{equation}
	\label{eq:AMC}
	M\sim \rho_{\rm mc}^{1/6}\,\frac{m_{\rm Pl}}{\mu} \,M_{\rm mc}^{1/3}\,.
\end{equation}
The accretion of axion dark matter can proceed during radiation-domination around BHs formed primordially, leading to a dense axion minihalo in which axion stars can form through gravitational cooling before galaxy formation~\cite{Hertzberg:2020hsz}. The timescale at which this condensation occurs once both gravity and self-interactions are taken into account is currently under debate~\cite{Kirkpatrick:2020fwd, Chen:2020cef, Chen:2021oot}.

\section{Oscillations and stability}
\label{sec:stability}

\subsection{Complex scalar fields}
The equilibrium configurations of BSs studied in the previous sections can be spoiled when a small perturbation is applied. Different BSs solutions can be classified as either belonging to the stable (S) or unstable (U) branch, depending on their behavior against perturbations~\cite{Lee:1988av, Gleiser:1988rq, Gleiser:1988ih, Jetzer:1988vr, Jetzer:1989qp, Jetzer:1989vs}. We first consider the mini-boson stars discussed in Sec.~\ref{sec:miniBS}, whose configuration is parametrized in terms of the field value at the core $\phi_c$. Under a small perturbation, a mini-boson star in the S branch oscillates and loses mass through the emission of relativistic bosons, to then relax in a stable configuration of smaller mass. Stars in the U branch would either collapse to form a BH if accreted, or migrate to a stable configuration through wave emission~\cite{Seidel:1990jh}. The critical value $\bar\phi_c$ that separates the solutions that are stable against small perturbations with $\phi_c \leq \bar\phi_c$ (S branch) from the unstable ones with $\phi_c > \bar\phi_c$ (U branch) can be found as~\cite{Lee:1988av, Gleiser:1988rq, Gleiser:1988ih}
\begin{equation}
	\frac{\mathrm{d}M}{\mathrm{d}\phi_c} = 0\,,\quad \frac{\mathrm{d}N}{\mathrm{d}\phi_c} = 0\,,
\end{equation}
where $N$ is the Noether charge in Eq.~\eqref{eq:noethercharge}. It turns out that the critical value $\bar\phi_c$ corresponds to the maximal mass configuration (the Kaup mass), so that ground state solutions of low mass $M < M_{\rm Kaup}$ are stable against perturbations~\cite{Gleiser:1988ih, Seidel:1990jh}. Generally, excited states are also unstable unless the conservation of the Noether charge is imposed~\cite{Lee:1988av, Jetzer:1989qp}. Massive boson stars show a similar pattern, with configurations of mass below the maximal mass in Eq.~\eqref{eq:colpilimit} belonging to the S branch that oscillate in response to small perturbations to settle onto new stable configurations. Massive stars would either collapse to a BH or lose mass and settle to a stable configuration when perturbed~\cite{Balakrishna:1997ej}. Refined computations show that BSs might have two stable branches, the first corresponding to the Newtonian configurations of mass below the maximal mass $M_{\rm max}$, together with denser relativistic configurations~\cite{Kleihaus:2011sx, Herdeiro:2021lwl}.

The equations for the perturbations are obtained by perturbing the field and the metric around the equilibrium solution by small quantities $\delta\phi$ and $\delta g_{\alpha\beta}$, respectively, such that the total number of particles $N$ is conserved. For scalar BSs with spherical symmetry and in the formalism of Sec.~\ref{eq:decomposing}, this corresponds to the choice~\cite{Gleiser:1988rq, Gleiser:1988ih, Jetzer:1988vr}
\begin{equation}
	u = u_0 +\delta u; \quad v = v_0 +\delta v; \quad \phi = \phi_0[1+ \delta\phi_R +i\delta\phi_I]\,,
\end{equation}
where the subscript ``0'' labels the background quantities and where perturbations are function of both $r$ and $t$. The resulting time-dependent equations for the perturbations depend on the radial oscillation frequency squared $\omega^2$, whose sign is related to the stability of the system. An alternative approach which allows the inclusion of both linear and non-linear perturbations consists in numerically solving EKG~\cite{Hawley:2000dt}. Recent 3D numerical simulations of the full relativistic expressions show that for both azimuthal quantum numbers $m = 1$ and $m = 2$, a mini-boson star shows instabilities against linear, non-axisymmetric perturbations~\cite{Siemonsen:2020hcg}.

Boson stars theories in which the complex field possesses a U(1) symmetry can be generalized by considering a U($\mathbb{N}$) symmetry field for an arbitrary odd value of $\mathbb{N}$. These states, called $\ell$-boson stars in which $\ell = (\mathbb{N}-1)/2$, are generally more massive and compact than ordinary BSs~\cite{Alcubierre:2018ahf}. The stability of $\ell$-boson stars is qualitatively similar to the corresponding results for mini-boson stars, with the maximal mass separating the family of $\ell$-boson stars in the S branch from those in the U branch~\cite{Alcubierre:2021mvs}.

A process that could potentially brings a BS to collapse into a BH corresponds to the interaction of a BS with an incident massless real scalar field which transfers energy through gravity~\cite{Brady:1997fj, Hawley:2000dt}. For this mechanism, at least two different processes can be loosely defined. For Type I processes, only BHs above a certain mass can be formed, while Type II processes could lead to BHs of any masses. The outcome of the process depends on the mass of the BS and on the wavelength of the incoming radiation compared to the Compton wavelength of the boson $\mu^{-1}$. Numerical simulations of mini-boson stars of nearly maximal Kaup mass show that the star compresses to then either dissipate or collapse to a BH, depending on the initial conditions~\cite{Hawley:2000dt}.

\subsection{Real scalar fields}

We now turn the attention to oscillatons and oscillons, which naturally tend to dissipate since they do not possess a conserved charge. Oscillons are not exact periodic solutions of the corresponding field equations and they decay in a finite time through classical and quantum processes. However, in the non-relativistic limit the wave function $\psi$ possesses and approximated U(1) symmetry which leads to an approximately conserved charge that is responsible for the longevity of these structures~\cite{Kasuya:2002zs, Mukaida:2016hwd},
\begin{equation}
	Q = \frac{i}{4}\int \mathrm{d}^3{\bf x}\left[\psi^*\dot\psi - (\dot\psi^*)\psi\right]\,.
\end{equation}
The effects of gravity can be similarly implemented by constructing an effective field theory in the non-relativistic regime~\cite{Eby:2018ufi}, and different schemes to include relativistic corrections have been implemented~\cite{Mukaida:2016hwd, Namjoo:2017nia, Braaten:2018lmj, Salehian:2021khb}.

The exact quantification of the lifetime is still under debate~\cite{Page:2003rd, Fodor:2009kg, Grandclement:2011wz, Degollado:2018ypf, Zhang:2020ntm, Kawasaki:2020jnw, Cyncynates:2021rtf}. The decay through quantized oscillons can dominate over the emission of classical radiation, with the outgoing radiation growing linearly in theories with a single field~\cite{Hertzberg:2010yz}. The properties of oscillatons can be assessed by numerically evolving EKG for the scalar field, so that the family of solutions divides into two sets. For a quadratic potential, a stable S branch oscillaton evolves according to the fundamental mode $\omega$ and is stable against small perturbations, while it removes a portion of its mass and migrates to a stable configuration of smaller mass in response to larger perturbations. On the other hand, an unstable U branch oscillaton migrates to the S branch even in response of a tiny perturbation, or collapses into a BH once the mass is increased. The transition between the two branches occurs at the critical mass $M_c = 0.605\,m_{\rm Pl}^2/\mu$ of compactness $\mathrm{C} = 0.14$~\cite{Alcubierre:2003sx}. Similar work on oscillons with a quartic self-interaction has revealed the existence of similar S branches~\cite{Valdez-Alvarado:2011onf}. The lowest radial oscillation modes of an oscillaton within a theory with a repulsive self-interaction have been recently computed with this method~\cite{Lopes:2019eue, Kain:2021rmk}.

In string motivated scenario, an oscillon that would disperse in the absence of gravity would either turn into a metastable configuration or collapse to a BH when gravity is included; a metastable oscillon in the absence of gravity would be brought to collapse when gravity is included, depending on the mass and self-interaction model considered~\cite{Muia:2019coe}. Using the open source numerical code $\textsc{GRChombo}$~\cite{Clough:2015sqa}, a similar analysis covering a larger class of potentials leads to the mapping of the region of parameter space in which the star would be brought to collapse, disperse, or would remain metastable whether gravity is included or not~\cite{Nazari:2020fmk}.

The stability of axion stars has received particular attention in the literature. The collapse of an axion star to form a BH has been studied in the classical limit by solving numerically EKG, revealing the existence of a stability diagram parametrized by the mass of the axion star $M$ and the axion decay constant $f_a$~\cite{Helfer:2016ljl}. Depending on the values of $(M, f_a)$, three different regions are identified corresponding to long-lived axion star solutions, collapse to a BH, and complete dispersion, with the boundaries meeting at a triple point $(M, f_a) \sim (2.4\,M_{\rm Pl}^2/\mu, 0.3\,M_{\rm Pl})$. Following the collapse numerically reveals the existence of a ``bosenova'' explosion corresponding to an emission of relativistic axions in subsequent bursts; the stable remnant would collapse to a BH only for $f_a$ close to the Planck scale~\cite{Levkov:2016rkk, Eby:2016cnq, Eby:2017xrr, Michel:2018nzt}. Quantum effects can also destabilize the star, allowing for the axion field $\phi$ to decay through processes such as $3\phi\to \phi$. This process is not relevant for the ``dilute'' axion stars studied in Sec.~\ref{sec:axionstars}, whose lifetime exceeds cosmological timescales~\cite{Eby:2015hyx}.

Compact axion stars can decay into radio photons through parametric instability, when conditions are met so that an axion decaying into photons stimulates the additional decay of more axions in a cascade that amplifies exponentially the number of photons~\cite{Tkachev:1987cd, Riotto:2000kh}. This effect has been studied in dense clumps of axions~\cite{Tkachev:2014dpa, Hertzberg:2018zte, Arza:2018dcy, Carenza:2019vzg}, including axion stars which could contribute to both the intergalactic radio background and fast radio bursts~\cite{Levkov:2020txo}.

Although feeble, a coupling between the axion and the photon arises within most axion model through loops involving charged fermions. This provides a channel that allows an axion star to dissipate its energy in the magnetized medium surrounding white dwarfs and NSs~\cite{Bai:2017feq, Buckley:2020fmh, Bai:2021nrs}.\footnote{\,The same is expected to occur for axion miniclusters encountering NSs~\cite{Edwards:2020afl, Xiao:2021nkb}.} Results greatly differ depending on the maximal value of the axion star mass that can be attained, ranging from optimistic values using $M_{\rm max} \sim 10^{-5}\,M_\odot$~\cite{Iwazaki:1999ub, Iwazaki:1999my} to more conservative values using $M_{\rm max} \sim 10^{-14}\,M_\odot$~\cite{Barranco:2012ur}, assuming an axion of mass $\mu = 10^{-5}\,$eV. The value quoted in Eq.~\eqref{eq:maxaxionstarmass} is somewhat intermediate between these results. However, the actual computation requires input from the mass distribution of axion stars in the Galaxy, the so-called halo mass function, which is currently an unknown feature for these objects and a limit even for the assessment of the minicluster distribution~\cite{Kavanagh:2020gcy}.

\section{Boson stars and gravitational waves}
\label{sec:GW}

Gravitational waves (GWs) have opened a new window for probing the strong-field regime of general relativity (GR). The array of detectors that includes Advanced LIGO~\cite{LIGOScientific:2014pky}, Advanced Virgo~\cite{VIRGO:2014yos}, KAGRA~\cite{Aso:2013eba}, GEO600~\cite{Grote:2010zz}, with the addition of the upcoming LIGO-India~\cite{Fairhurst:2012tf}, are detecting astrophysical phenomena such as the coalescence of compact objects locked in binary systems. Generally, the coalescence of a binary system consists of three distinct phases. In the early inspiral phase, the objects are sufficiently apart so that the post-Newtonian approximation suffices~\cite{Blanchet:2013haa}. The merger phase describing the collision between the two compact objects requires a treatment in terms of numerical relativity~\cite{Font:2008fka, Faber:2012rw}. Finally, in the ringdown phase the resulting object relaxes to equilibrium, emitting GWs in the process~\cite{Buonanno:1998gg, Kokkotas:1999bd}.

\begin{figure}
	\includegraphics[width = 0.8\linewidth]{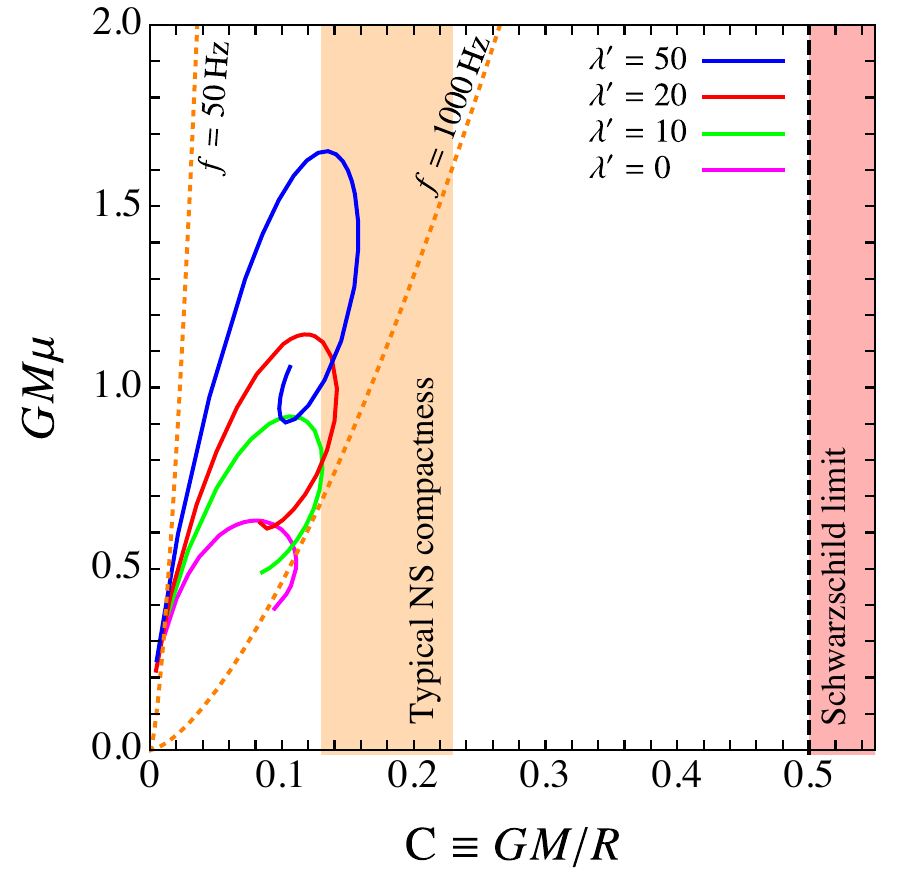} 
	\caption{Mass of a BS as a function of the compactness for the self-interaction parameter $\lambda' = 0$ (magenta), $\lambda' = 10$ (green), $\lambda' = 20$ (red), and $\lambda' = 50$ (blue). The light orange band shows the values of the typical compactness for neutron stars, while the light red region marks the region where the compactness of an object would be larger than of a Schwarzschild BH. The region bound between the two dotted orange lines falls within the detectability of the LIGO interferometers $50{\rm \,Hz} \lesssim f \lesssim 1000\,$Hz for the boson mass $\mu = 10^{-10}\,$eV, see Eq.~\eqref{eq:fISCO}.}
	\label{fig:compactness}
\end{figure}

Besides compact objects that are predicted from SM physics, such as astrophysical BHs and NSs, other exotic objects such as BSs could possess the required compactness to generate a GW strain that is detectable with present or near-future technologies. Consider two compact objects of similar mass and size and each of compactness $\mathrm{C}$, forming a system of total mass $M_{\rm tot}$. The frequency of the emitted GW spectrum at the end of the inspiral phase, when the stars occupy the innermost stable circular orbit, is~\cite{Giudice:2016zpa}
\begin{equation}
	\label{eq:fISCO}
	f = \frac{\mathrm{C}^{3/2}}{3\sqrt{3}\pi G M_{\rm tot}}\,.
\end{equation}
for example, an axion star is detectable by the LIGO/VIRGO network for an axion mass $10^{-9}{\rm \,eV} <\mu < 10^{-11}\,$eV~\cite{Widdicombe:2018oeo}. In Fig.~\ref{fig:compactness} we plot the mass of a BS as a function of its compactness for the self-interaction parameter $\lambda' = 0$ (magenta), $\lambda' = 10$ (green), $\lambda' = 20$ (red), and $\lambda' = 50$ (blue). The dotted orange lines in Fig.~\ref{fig:compactness} bound the region in which the frequency of the GW strain when the inspiral phase ceases, Eq.~\eqref{eq:fISCO}, lies within the LIGO detectability bandwidth, $50{\rm \,Hz} \lesssim f \lesssim 1000\,$Hz~\cite{LIGOScientific:2014pky}, for the mass of the boson $\mu = 10^{-10}\,$eV. The compactness of BSs is compared with that of a Schwarzschild BH $\mathrm{C}_{\rm BH} = 1/2$ and of neutron stars for which, using realistic assumptions for the equation of state, the compactness is expected to be $0.13 \lesssim \mathrm{C}\lesssim 0.23$~\cite{Lattimer:2006xb}.

Massive BSs can achieve a compactness similar to that of neutron stars and are ideal exotic candidates to be searched through GW interferometry~\cite{Croon:2018ybs}. Compact clumps of boson particles in binary configurations can mimic the GW emission from binary black hole mergers, depending on the details of the scalar self-interaction~\cite{Cardoso:2016rao}. Massive vector bosons form Proca stars (see Sec.~\ref{sec:procastar}), extremely compact object which are stable against non-axisymmetric perturbations, see Sec.~\ref{sec:procastar}. Rotating Proca stars can mimic some properties of Kerr BHs~\cite{Brito:2015pxa, Herdeiro:2021lwl} and are exotic candidates to explain the observation of the event GW190521 by the Advanced LIGO/Virgo detectors~\cite{Bustillo:2020syj}. A quartic coupling yields the maximum compactness $\mathrm{C}_{\rm max} \approx 0.16$ regardless of the coupling~\cite{Amaro-Seoane:2010pks}. Nontrivial self-interactions of light singlet scalars such as a logarithmic or Liouville potential could lead to BSs with a compactness similar to that of NSs, which could form binary systems whose GW strain would lead to a signal to noise ratio that is high enough to be detected by the LIGO-VIRGO collaboration~\cite{Choi:2019mva}. When a V-shaped scalar field potential is considered, the resulting BSs are extremely compact and stable below the maximal value of the mass, at which the compactness is similar to that of a Schwarzschild BH~\cite{Hartmann:2012da}.

Compact BS binaries could be easily within reach of next-generation GW detectors such as LISA~\cite{Sennett:2017etc, DiGiovanni:2020ror, Toubiana:2020lzd}, and the stochastic background of GWs from BS binaries could be detected in EPTA and LISA~\cite{Croon:2018ftb}. LISA could also reveal the existence of ultralight bosons through the superradiant instabilities induced in spinning BHs~\cite{Brito:2017zvb}. If the boson field that makes up the BS does not interact with ordinary matter through any other potential other than gravity, an exotic ``dark'' BS would form. A system of two binary dark BSs can be clearly distinguished from the merging of a binary system of two massive BSs with a self-interacting potential~\cite{Bezares:2017mzk, Palenzuela:2017kcg} from the GW signature and for different values of the compactness~\cite{Bezares:2018qwa}. Another possibility is to consider the emission of GWs from of extreme mass ratio inspirals composed of a BS and a supermassive BH, which can be distinguished from other binary systems because of the effects of tidal deformation and stripping~\cite{Guo:2019sns}.

GW searches are suitable not only for the search of exotic compact objects such as BSs, but also to probe their internal density profile. The coalescence of a binary system of two BSs with no event horizon but with a light ring can mimic the initial part of the post-merger ringdown phase from two coalescing BHs, while the later stages of the ringdown phase could reveal the nature of the compact object through GW echoes~\cite{Cardoso:2016oxy, Cardoso:2017cqb, Cardoso:2019rvt, Maggio:2021ans}. When a light ring forms, the object is sufficiently compact to mimic the vibration modes of a BH~\cite{Cunha:2015yba}. However, a BS becomes sufficiently compact in a region of parameter space in which it is also unstable against perturbations, making it difficult to build a BS without an event horizon but with a light ring which could fully mimic a BH~\cite{Cunha:2017wao}.

One issue for this type of search is related with the templates used for recognizing a detected signal as due to the coalescence of two compact objects. As discussed by the LIGO-VIRGO collaboration~\cite{LIGOScientific:2020tif}, template banks encoding binary black hole mergers are used to search for the gravitational waves emitted during the coalescence. For small deviations of the waveform, current template banks can be used to detect exotic compact objects~\cite{Krishnendu:2017shb, Kastha:2018bcr}, however such templates fail at reliably searching for objects with large quadrupoles such as those which possess a high spin or have a large mass split with the binary companion~\cite{Chia:2020psj} or Kerr BHs with scalar hair~\cite{Cunha:2015yba}. More simulations to create a database of adequate templates is desired.

GW searches can be adapted to look for exotic self-gravitating stellar objects such as BSs by reconstructing the values of the lowest multipole moments of each inspiraling object in the binary system, such as its moment of inertia $I$ and quadrupole moment $Q$. Further information can be obtained by measuring the tidal Love number $k_2$, which accounts for the response of a self-gravitating object to the tidal perturbations of an external field~\cite{Mora:2003wt} and carries information about the internal structure of a compact object. For example, the assessment of the Love number allows for a clean determination of the properties of a neutron star from the early inspiral phase~\cite{Flanagan:2007ix, Hinderer:2007mb, Hinderer:2016eia}. On the other hand, a consequence of the BH no-hair theorem is that the Love number of a BH is exactly zero, as can be shown with an explicit computation of the effects of tidal perturbation of a Schwarzschild BH~\cite{Fang:2005qq, Binnington:2009bb, Damour:2009vw} and to a Kerr BH. The tidal Love number has been calculated in the full relativistic approach for the lowest moments of a mini-boson star~\cite{Mendes:2016vdr}, for BSs in the presence of self-interacting scalar fields and to higher-multipole~\cite{Cardoso:2017cfl, Sennett:2017etc}, and for Proca stars~\cite{Herdeiro:2020kba}. Useful relations between the Love number, the moment of inerzia, and the quadrupole moment can potentially be used to efficiently combine multiple observations and assess the nature of the compact object~\cite{Maselli:2017vfi}. A dedicated analysis of the post-merger phase accessible in future detectors will also allow to probe the nature of boson-fermion star inspirals, see Sec.~\ref{sec:BFS}, and possibly discriminate between a neutron star and a degenerate object of neutrons admixed with bosons~\cite{Bezares:2019jcb}. Along with the post-merger properties of inspirals, the presence of bosonic DM would also affect the maximum mass and tidal deformability of a fermion degenerate object~\cite{Karkevandi:2021ygv}.

\section{Boson stars and dark matter}
\label{sec:DM}

The existence of non-luminous (dark) matter has been postulated to explain the observations of galactic rotation curves in regions of the halo extending much farther than the luminous component of a galaxy~\cite{1976AJ.....81..687R, 1976AJ.....81..719R}. This dark component is usually assumed to consist of massive particles with very low thermal velocities and non-relativistic (Cold Dark Matter, CDM~\cite{Peebles:1982ff, Bond:1982uy, Blumenthal:1982mv, Padmanabhan:2002ji, Peebles:2002gy}). Work on colliding galaxy clusters seem to confirm the existence of dark matter dominating the mass content of spiral galaxies and galaxy clusters~\cite{Clowe:2006eq, Bradac:2008eu}. An indirect confirmation of this picture comes from the success of the concordance cosmological model, or $\Lambda$ Cold Dark Matter ($\Lambda$CDM) model, in reproducing the anisotropies observed in the cosmic microwave background (CMB)~\cite{Liddle:1993fq}. Among the most promising candidates for the CDM component are the Weakly Interacting Massive Particle or WIMP~\cite{Bertone:2004pz}, or a population of zero-momentum axions~\cite{Preskill:1982cy, Abbott:1982af, Dine:1982ah, Stecker:1982ws, Hertzberg:2008wr, Visinelli:2009zm}. See Refs.~[\citen{Lisanti:2016jxe, Freese:2017idy}] for reviews.

However, several problems in reproducing observed properties of galaxies seem to lurk within CDM models, most remarkably the overabundance of small scale structure known as the ``missing satellite'' problem~\cite{Kauffmann:1993gv, Klypin:1999uc, Moore:1999nt}, the presence of a central density cusp or the ``cusp'' problem~\cite{Navarro:1996gj, Moore:1999gc, Ostriker:2003qj, Romanowsky:2003qv}, and too many massive dense subhalos compared with satellites around the Milky Way  or the ``too big to fail'' problem~\cite{Boylan-Kolchin:2011qkt}). In more detail, observations of both nearby dwarf galaxies and low surface brightness galaxies show that the density profile of the CDM halo at the core reaches a constant value~\cite{Moore:1994yx, Burkert:1995yz}. In contrast, various N-body simulations predict that the CDM density distribution steepens at the center of the halo~\cite{Navarro:1994hi, Navarro:1995iw, Navarro:1996gj, Fukushige:1996nr, Moore:1997sg, Ghigna:1998vn}. To address these discrepancies, models that step away from CDM and introduce additional properties to the DM are usually considered.

Among the solutions proposed to explain the dark matter puzzle discussed, it has been suggested that dark matter could consist of a coherent scalar field with long range correlation, whose quanta are very light particles~\cite{Sin:1992bg, Ji:1994xh, Sahni:1999qe, Lee:1995af, Hu:2000ke, Peebles:2000yy}. In fact, on short length scales, light scalar fields do not behave as CDM and quantum pressure would inhibit cosmological structure growth~\cite{Marsh:2013ywa, Hlozek:2014lca}. This model suppresses low mass galaxies and provide cored profiles in CDM-dominated galaxies~\cite{Hu:2000ke, Peebles:2000yy, Marsh:2015wka}. See Ref.~[\citen{Hui:2021tkt}] for a review.

Consider for example a mini-boson star of Kaup mass as in Eq.~\eqref{eq:kauplimit}. The weight of such an object would be similar to that of a galaxy, $M_{\rm gal} \sim 10^{12}\,M_\odot$, for the mass of the boson $\mu \sim 10^{-22}\,$eV. A complex scalar field with a self-interaction has also been considered to explain the large-scale structure of galaxies~\cite{Press:1989id}, even in relation with the formation of a massive BS~\cite{Lee:1995af}. The cosmology of complex scalar fields has also been assessed in the presence of self-interactions, which allows for a better fit with data from CMB and big bang nucleosynthesis~\cite{Li:2013nal, Li:2016mmc}. Real scalar fields with a cosh-type self-interaction potential could also form oscillons of galactic mass $M_{\rm gal}$~\cite{Alcubierre:2001ea}. Adaptively refined simulations of the Schr{\"o}dinger-Poisson equations resolving the gravitational collapse of wavelike DM with a mass of the order of $10^{-22}\,$eV show the effects on structure formation due to their large Jeans length~\cite{Schive:2014dra, Schive:2014hza, Schive:2015kza}, which improved over previous work on the subject~\cite{Goodman:2000tg, Riotto:2000kh, Boehmer:2007um}. The high resolution achieved in recent simulations with wavelike DM show that the model reproduces the behaviour of CDM at large scales, while the behavior departs great for lengths smaller than the Compton wavelength of the boson where a BS forms at the core of dwarf galaxies with the halo virial mass $M_h$ and the solitonic core mass $M_c$ being related as $M_c \propto M_h^{1/3}$~\cite{Schive:2014dra, Schive:2014hza, Veltmaat:2018dfz, Bar:2018acw, Chavanis:2019faf}, see Eq.~\eqref{eq:AMC}. Recent simulations of structure formation with wavelike DM at high-redshift, including the effects of baryons, highlight the observable differences with the CDM scenario such as reduced stellar formation rate in the flattened core of galaxies and the formation of dense DM filaments accessible to the James Webb Space Telescope~\cite{Mocz:2019pyf}.

If dark matter is in bosonic particles, its presence could be inferred from neutron stars in binary configurations, since dark matter accretion around them could lead to an observable peak in the GW spectrum that differs from the features induced by the neutron components~\cite{Ellis:2017jgp, Ellis:2018bkr}. A similar effect occurs for accretion around binary black holes regardless of the particle nature of DM~\cite{Kavanagh:2020cfn}.

Bosonic DM could even form a Bose-Einstein Condensate (BEC), described by the Gross-Pitaevskii or non-linear Schr{\"o}dinger equation, see Sec.~\ref{sec:BEC} and Ref.~[\citen{Suarez:2013iw}] for a general review of the models proposed. One example is the axion, whose galactic condensation can be modeled as a coherent BEC with small spatial gradient~\cite{Sikivie:2009qn, Park:2012ru, Arias:2012az, Hlozek:2014lca}. Data for the extensions of galaxies and galaxy clusters can be used to constrain the mass and the interaction strength of spin-zero and spin-one particles~\cite{Pires:2012yr, Banik:2013rxa}. The cosmological evolution of a BEC DM component has also been extensively explored~\cite{Ferrer:2004xj, Grifols:2005kv, Fukuyama:2007sx, Fukuyama:2009vzr}. The self-interacting BEC would exert a dynamical friction on the massive objects that move through the medium such as stars, planets, or BHs, whose effect can be explored both via linear perturbation theory and with numerical simulations~\cite{Annulli:2020lyc, Annulli:2020ilw}. The matching with the observed orbital decay times due to dynamical friction of some of the globular clusters in Fornax help the assessment of the properties of such a BEC~\cite{Hartman:2020fbg}.

\section{Other searches}
\label{sec:othersearches}

\subsection{Lensing}

The magnitude of the deflection of light from a massive body acting as a lens has been the smoking gun allowing to distinguish GR from Netwonian physics, since the former predicts twice the deflection angle than the latter. Lensing from compact objects such as BSs could cause a measurable change in the brightness of the lensed objects, an effect known as microlensing~\cite{Paczynski:1985jf}. A photon of angular momentum $\ell$ and total energy $\epsilon$ lensed from a compact object such as a BH or a NS travels along the null geodesics of the metric in Eq.~\eqref{eq:metric} with $\vartheta=\pi/2$, given by~\cite{Misner:1973prb}
\begin{eqnarray}
	\left(\frac{\mathrm{d}r}{\mathrm{d}\tau}\right)^2 + e^{-u}\frac{\ell^2}{r^2} &=& \epsilon^2e^{-u-v}\,,\\
	e^v \frac{\mathrm{d}t}{\mathrm{d}\tau} &=& \epsilon\,,\\
	r^2 \frac{\mathrm{d}\varphi}{\mathrm{d}\tau} &=& \ell\,,
\end{eqnarray}
where $\tau$ is the affine parameter along the null geodesics. For a mini-boson star of Kaup mass, the lensing typically produces three images, two of them being inside the Einstein radius and one outside of it~\cite{Dabrowski:1998ac}. Microlensing effects from BSs with generic self-interactions have also been explored~\cite{Schunck:1999zu}.

Surveys of microlensing events pointing towards various sources such as the Magellanic Clouds (EROS-2) and the Galactic centre (OGLE-IV) can help constraining the population of BSs in the Milky Way. It turns out that the fraction of DM which is in boson stars should amount to less than $\mathcal{O}(10^{-2})$ for masses $M \in [10^{-6}, 10^{-1}]\,M_\odot$ and for point-like BSs, the exact value of the bound depending on the survey considered and on the radius containing 90\% of the lens mass~\cite{Croon:2020wpr}. Accounting for the finite size of the BS is indeed important when their radius is larger than about $0.1\,R_\odot$~\cite{Croon:2020ouk}.

An interesting phenomenon occurs when considering lensing from an axion star, where axion-photon interactions can bend light rays even more efficiently than gravitational deflection. Such an effect is in range of the Square Kilometer Array, which could detect axion stars in the mass range $M \in [10^{-14}, 10^{-11}]\,M_\odot$ through the anomalous shifts in the apparent positions of background radio sources~\cite{Prabhu:2020pzm}. It has been recently claimed that axion stars could be responsible for some microlensing candidate events reported by the OGLE-IV and Subaru Hyper Suprime-Cam, in a region of space where these stars have a mass of the order of the Earth and make a substantial ($\gtrsim 10\%$) fraction of the DM~\cite{Sugiyama:2021xqg}. This claim has been challenged in the context of the QCD axion, where such heavy masses for the axion stars can be reached only if the axion is relatively light, see Eq.~\eqref{eq:maxaxionstarmass}, and the PQ SSB occurs during the inflation period~\cite{Schiappacasse:2021zlr}. The finite size of the source and the lens can also affect the outcome of the search~\cite{Katz:2018zrn, Fujikura:2021omw}.

\subsection{Imaging}

Material accreting around a rotating BH would heat up and emit in a range of frequencies, revealing the existence of a dark shadow in combination with a bright emission ring. This image is within reach of very long baseline interferometry experiments such as the Event Horizon Telescope (EHT), a global network of radio telescopes observing at $1.3\,$mm wavelength~\cite{galaxies4040054}. The EHT collaboration has recently imaged the shadow of the supermassive BH residing at the centre of the giant elliptical galaxy Messier 87, M87$^*$~\cite{EventHorizonTelescope:2019dse}, and will soon produce the first image of the supermassive BH at the center of the Galaxy, Sagittarius A$^*$ (SgrA$^*$).

Can supermassive and rotating BSs mimic this? The mass distribution of a BS extends to infinity, so that these compact objects do not possess an event horizon or even a defined surface. Nevertheless, a BS could also be surrounded by the accretion flow of a rotating ring of plasma similarly to an accreting BH. Preliminary work simulating the image produced at $1.3\,$mm wavelength from an accreting BS mimicking SgrA$^*$ showed such a similarity~\cite{Vincent:2015xta, Vincent:2016sjq}. The accretion ring around SgrA$^*$ seems to reproduce the results from the BH counterpart, including the size of the shadow for given mass and spin. It is also expected that a BS would accrete at a much slower rate with respect to the BH counterpart, which is consistent with the observations at our Galactic centre~\cite{Torres:2000dw, Troitsky:2015mda}.

Magnetohydrodynamic simulations in the GR framework of accreting non-rotating mini-boson stars show that these objects can be distinguished from both non-rotating and moderately spinning BHs when observed with a realistic setup such as that of EHT, thanks to the differences related to the absence of an event horizon for the BS~\cite{Olivares:2018abq}. The inclusion of an orbiting space antenna on an elliptical orbit to the baseline would considerably increase the angular resolution of EHT. This new setup would further increase the capability of EHT to distinguish a rapidly spinning Kerr BH from a BS~\cite{Fromm:2021flr}.

\subsection{Collision of boson stars and oscillatons}

Fully relativistic numerical simulations of head-on collisions of BSs and oscillatons lead to different results depending on their relative phase difference. In fact, the ansatz for a pair of non-rotating BSs can be taken as~\cite{Palenzuela:2006wp}
\begin{equation}
	\Phi_1=\phi_1(r)e^{-i\omega t}\,,\qquad \Phi_2=\phi_2(r)e^{-i\epsilon\omega t +\theta}\,,
\end{equation}
where $\theta$ is the relative phase difference and $\epsilon = \pm 1$ gives the sign of the Noether charge. The second star can be an in-phase BS $(\epsilon = 1, \theta = 0)$, a BS in opposition of phase $(\epsilon = 1, \theta = \pi)$ or an anti-BS $(\epsilon = -1, \theta = 0)$. Energy considerations lead to the total energy density for the system $\rho_{\rm tot} = \rho_1 + \rho_2 + \Delta_0\cos[(1-\epsilon)\omega t-\theta]$, where $\Delta_0$ is a positive-definite quantity~\cite{Palenzuela:2006wp}. A head-on collision between two mini-boson stars would have $\rho_{\rm tot} > \rho_1+\rho_2$, so that the merging of the two stars is favored. If the second BS is in opposition of phase, $\rho_{\rm tot} < \rho_1+\rho_2$ the merging does not occur. Finally, if the second object is an anti-BS, $\rho_{\rm tot}$ has a time varying component and the outcome of the collision depends on the relative magnitude between the oscillation timescale $1/\omega$ and the interaction time scale. When a repulsive interaction is present a similar result is found, in particular when the stars are in opposition of phase they would first bounce and lose kinetic energy before getting at rest without merging~\cite{Cardoso:2016oxy}.

A similar mechanism is observed for in-phase oscillaton mergers, where the outcome of the collision and the spectrum of emitted GW radiation depends on the compactness~\cite{Helfer:2018vtq, Helfer:2021brt}. Depending on the phase, the collision of two massive solitons would rapidly proceed to form a BH and the event could mimic that of two colliding BHs, while the modes of massive oscillatons would have higher amplitudes that could make the distinction easier.

A collision of two ultra-relativistic BSs could lead to the formation of a BH. This result is related to Thorne?s hoop conjecture, stating that compressing an amount of energy $E$ into a spherical region of radius below the Schwarzschild radius $R_s \equiv 2GE$~\cite{1972mwm..book..231T}. Applying this conjecture to a pair of colliding BSs, each of mass $M$, radius $R$ and relative speed $v$ in the center of mass, leads to the prediction that a BH forms when the compactness exceeds $\mathrm{C} > 1/(4\gamma)$, where $\gamma = (1-v^2)^{-1/2}$. Numerical computations show that a BH is formed from this process~\cite{Rezzolla:2012nr}, even at more modest relative speeds $\gamma \gtrsim 1/(12\mathrm{C})$~\cite{Choptuik:2009ww, East:2012mb, Widdicombe:2019woy}. Head-on collisions of oscillatons also lead to BH formation if their compactnesses lie above $\mathrm{C} \gtrsim 0.035$, although a slight boost prevents BH formation even about this threshold~\cite{Widdicombe:2019woy}.

Using the numerical code $\textsc{GRChombo}$~\cite{Clough:2015sqa}, the collision of an axion star of a given compactness with a spinning BH leads to most of the axion star to be swallowed by the BH, except in a fraction of cases in which an axion cloud of mass $\mathcal{O}(0.1)$ that of the BH gets gravitationally trapped outside of the BH forming a dark cloud~\cite{Clough:2018exo}. The collision of an axion star with a NS of typical mass $\sim 1.38\,M_\odot$ would lead to the perturbation and the collapse of the axion star into a BH within the potential well of the NS, which could lead to many multi-messenger observables such as the emission of neutrinos and GWs as well as electromagnetic signals of various nature~\cite{Dietrich:2018jov}.

\section{Concluding remarks}

Boson stars and oscillatons have long been discussed in the literature in relation to a variety of phenomena, and are currently experiencing a resurgence because of the refined techniques that could verify their existence discussed in Secs.~\ref{sec:GW}-\ref{sec:othersearches}. The interest in the study of these non-topological solitons ranges from the purely mathematical one to their appearance in field theoretical models to their impact in astrophysics and cosmology. The importance of these studies is signaled by the amount of reviews and the rapid advances in the field, with the hope to claim a clean detection in the next decade.

The study of compact objects from boson fields is now in a mature stage as precision numerical results on the formation in astrophysical and cosmological settings exist, as well as various predictions on the interplay among these objects and other compact objects such as BHs and NSs. It is not unrealistic to believe that exotic objects could appear soon in the GW searches of coalescent compact objects, as discussed extensively in Sec.~\ref{sec:GW}. New physics could appear through indirect observations that involve bosonic compact objects and could even shed light on the post-inflation epoch. In this view, the collection of work reported in this review aims at boosting this line of search and favor the discussion in the community.

\section*{acknowledgments}
I would like to thank Pierre-Henri Chavanis for his useful suggestions that improved the quality of this review. I acknowledge support from the European Union's Horizon 2020 research and innovation programme under the Marie Sk{\l}odowska-Curie grant agreement No.~754496 (H2020-MSCA-COFUND-2016 FELLINI).

\newpage
\bibliographystyle{ws-ijmpd}
\bibliography{BosonStars}

\end{document}